\documentclass[letterpaper,twocolumn,10pt]{article}
\usepackage{usenix}

\usepackage{filecontents}
\usepackage{float}
\usepackage{mwe}
\usepackage[]{graphicx}
\usepackage{tikz}
\usepackage{amsmath}
\usepackage{multirow}
\usepackage{caption}
\usepackage{subcaption}
\usepackage[table,xcdraw]{xcolor}
\usepackage{mathtools}  
\usepackage{amssymb} 
\usepackage{float}
\usepackage{xspace}
\usepackage{tikz}
\usepackage{algorithm}
\usepackage{algpseudocode}
\usepackage{listings}
\usepackage[utf8]{inputenc}
\usepackage{fancyvrb} 
\usepackage{url}
\usepackage{cite}
\usepackage{placeins}

\usepackage{makecell}

\newcommand{\dupin}[0]{\texttt{\textbf{DUPIN}}\xspace}

\newcommand{\tikzcircle}[2][red,fill=red]{\tikz[baseline=-0.5ex]\draw[#1,radius=#2] (0,0) circle ;}%
\usepackage[dvipsnames]{xcolor}

\usepackage{xparse}
\usepackage{xcolor} 
\usepackage{todonotes} 

\usepackage[available]{usenixbadges}

\newcommand{\circnum}[1]{%
\tikz[baseline=(char.base)]{
\node[shape=circle,fill=black,text=white,
inner sep=1pt,font=\small] (char) {#1};}
}

\newcommand{\colortext}[2]{{\textcolor{#1}{\textbf{#2}}}}
\newcommand{\boldred}[1]{{\colortext{red}{\textbf{[#1]}}}}
\newcommand{\boldblue}[1]{\colortext{blue}{\textbf{[#1]}}}

\begin{document}

\pagestyle{empty}
\pagenumbering{gobble}

\date{}

\title{\Large \bf \dupin{}: Attack Learning Is Still Needed! \underline{D}emonstrating Few-Shot after \underline{U}nsupervised \underline{P}retraining \underline{I}s A \underline{N}imble Forensics Learner}

\author{
 {\rm Chanwoo Bae}\\
 Purdue University
 \and
 {\rm Hailun Ding}\\
 IBM Research
 \and
 {\rm Shiqing Ma} \\
 UMass Amherst
 \and
 {\rm Xiangyu Zhang}\\
 Purdue University
} 

\maketitle
\begin{abstract}
Advanced persistent threats (APTs) pose a significant challenge in cybersecurity, involving staged and prolonged operations that often remain undetected until postmortem indicators, such as sabotage or financial loss, emerge. 
In consequence, human analysts are facing a needle-in-a-haystack challenge among vast accumulation of daily audit logs. However, leveraging a learning system for attack forensics is limited due to the scarcity of attack data, as attacks occur infrequently.
In addition, since malicious behaviors are usually embedded in massive benign ones, they are very hard to label by humans.
Thus, the recent approaches leverage self-supervised learning methods, where models rely solely on benign data and perform outlier detection. However, these methods struggle with the increasing complexity and dynamics of large-scale audit logs, often resulting in non-trivial false positives. Therefore, we propose a novel approach to learning-based attack forensics called \dupin{}.
First, \dupin{} performs unsupervised pre-training on an enormous amount of audit events in the form of provenance graphs.
It then proceeds to a few-shot learning stage, leveraging a small number of labeled attack examples to fine-tune its detection capabilities.
We pretrain \dupin{} on up to 38 - 52 days of audit logs (7.3TB total) and evaluate it against various baselines on 25 APT campaigns across four different data sources, facilitating the scalable evaluation. 
\end{abstract}

\begin{figure*}[tb]
    \centering
    \begin{subfigure}[b]{0.26\textwidth}
        \includegraphics[width =\textwidth]{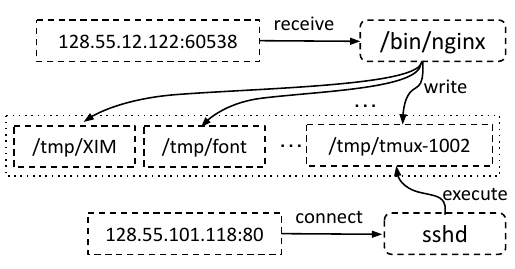}
        \caption{\centering \textit{Attack Example - Target}: Deploying multiple backdoor files.}
    \end{subfigure}%
    \begin{subfigure}[b]{0.30\textwidth}
        \includegraphics[width =\textwidth]{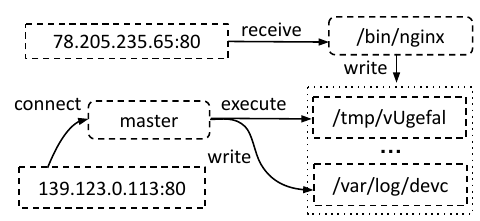}
        \caption{\centering \textit{Attack Example - Training}: Compromising the victim's PC.}
    \end{subfigure}%
    \begin{subfigure}[b]{0.36\textwidth}
        \includegraphics[width = \textwidth]{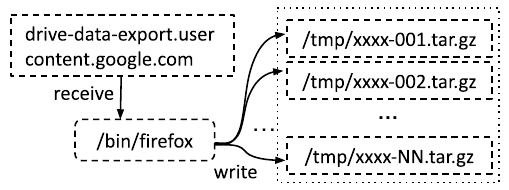}
        \caption{\centering \textit{Benign Example - Training}: Downloading fragmented \textit{.tar.gz} files from \textit{Google Drive}.}
    \end{subfigure}%
	\caption{Motivation example.}
	\label{fig:motivation-example}
\end{figure*}

\section{Introduction}\label{sec:intro}

Advanced persistent threats (APTs) are sophisticated cyberattacks, characterized by their extended timeframes and organizational/state-level targets. These meticulously planned operations unfold in stages, often remaining undetected for long periods. Victims typically recognize APTs only after suffering substantial damage, such as critical infrastructure sabotage or severe financial losses. Despite the significant harm APTs can inflict, their stealthiness and complexity make investigation challenging. The vast accumulation of daily audit logs turns surveillance into a needle-in-a-haystack challenge. Thus far, attack prevention has largely relied on human analysts, who must closely monitor extensive logs to identify a small set of potential threats. However, due to the limited supplies of trained experts, major challenges posed by complex attacks remain unsolved.

To reduce the complexity of analysis and human efforts, researchers have explored automation of auditing~\cite{inam2023sok}, leveraging learning-based methods for threat analysis~\cite{alsaheel2021atlas,zengy2022shadewatcher,yang2023prographer}. 
Specifically, these approaches reduce attack forensics to a classification problem, enabling the model to learn and make decisions based on data distributions.

However, a prominent challenge is that 
attacks appear sporadically and sparsely and hence attack data is lacking, making the traditional supervised model training difficult.
Consequently, the state-of-the-art methods employ self-supervised learning, such as one-class training, by primarily using benign data~\cite{ding2023airtag,jia2024magic}. The idea is to learn the distribution of the majority of data and label outliers as suspicious. Our evaluation (in Section~\ref{sec:eval-effectiveness}) 
shows that such methods have limited effectiveness as many malicious behaviors share substantial similarity with benign behaviors and unseen benign behaviors are often considered as malicious mistakenly. 

We consider that learning from attacks remains crucial. As evidenced by the recent rise of cyber threat intelligence~\cite{liao2016acing,shu2018threat}, attack knowledge has become a key factor in preventing and mitigating on-going cyberattacks. The primary insight of threat intelligence is that as attack actors become more advanced, their techniques are systematized, shared, and thus repeated. For example, after reprocessing the data from \textit{MITRE ATT\&CK} framework~\cite{mitre}, we discover that 85\% of the attack techniques are employed by two or more known threat actors, with each technique being used by an average of 18.6 actors (the number of real appearances can be even more). Inspired by these findings, we aim to develop a learning-based automated system, assuming attack training can operate with threat sharing and/or collaborative collection process. In this paper, we assume certain characteristics of cyberattacks and responses, where (i) attack sharing or collection provides an attack knowledge base in the form of raw logs or shareable models, even though the scale may be limited, and (ii) attack techniques are archived, shared among their actors, and thus would reappear.

Recently, the remarkable advancements in NLP techniques further support the use of attack-guided learning. For example,
the training phase can be divided into two steps where the model first undergoes extensive self-supervised pretraining~\cite{devlin2018bert,liu2019roberta}, and then addresses downstream tasks as classification problems. Consequently, such models can rapidly adapt to downstream tasks through few-shot learning or fine-tuning.
Audit data aligns perfectly with this 
learning paradigm. In particular, the majority of audit logs denote benign system behaviors and can be used in pretraining such that the model can understand system behaviors and provenance in general, like a foundation language model which understands natural languages. 
Then we can use the sparse attack samples for few-shot learning with the pretrained model.

We propose \dupin{}, a novel learning based forensic analysis system. 
It first uses an enormous repository of benign audit logs to pretrain a foundation model such that the model can understand the domain. As raw audit logs are very noisy, with interleaving events from numerous processes, we first transform them to provenance graphs and use them to pretrain a {\em Graph Attention Network} (GAT). 
We employ masked prediction tasks (i.e., node/edge prediction and classification) on provenance graphs in the unsupervised pretraining phase. 
After training, we further perform few-shot learning using a set of attack samples. 
In summary, we make the following contributions.

\begin{itemize}
  \itemsep0em 
  \item We propose a new approach to learning-based attack auditing, leveraging a foundation GAT and the model’s ability of rapid adaptation to the downstream task. In addition, we present the masked prediction learning on provenance graphs, enabling unsupervised pretraining. It is particularly suitable for the forensic domain.
  \smallskip
  \item We conduct extensive self-supervised learning on provenance graphs, using pretraining data spanning a total of 52 days of traces for {\tt Linux} and 38 days for {\tt Windows} with each tracepoint spanning 10 to 12 days. Our pretraining employs a self-supervised approach, utilizing masked learning on process and event names. The ablation study demonstrates that our approach achieves a 16\% improvement in accuracy as a result of extensive pretraining of the \dupin model.
  \smallskip
  \item We evaluate \dupin{} across 25 representative APT attack sets from four different dataset sources. We adopt a cross-validation strategy in which the model is restricted from learning from attack sets originating from the same red team (i.e., dataset source). In this setting, the training and validation sets are separated based on different attack dataset sources to more accurately reflect evolving, diverse conditions in real-world attack surveillance. 
  \smallskip  
  \item We implement and release a toolkit to facilitate the log forensics. It is compatible with three audit formats: (i) \texttt{Linux Audit Framework}, (ii) \texttt{Windows ETW}, allowing to directly collect logs from \texttt{Ubuntu}/\texttt{Windows} environment for simulation purposes, and (iii) \texttt{Common Data Model} (CDM), covering all tracepoints in \textit{DARPA TC} dataset. The set of APIs supports raw log parsing and graph exploration. This toolkit is designed to facilitate future research and enable comparisons of different approaches on a common foundation to prevent environmental differences in implementation and tedious data process. We release an repository of our toolkit.\footnote{\url{https://doi.org/10.5281/zenodo.20598994}} The implementation of \dupin is included in the repository.
 
\end{itemize}
\begin{figure*}[tb]
    \centering
    \vspace{3mm}
    {%
    \setlength{\fboxsep}{2pt}%
    \setlength{\fboxrule}{0.5pt}%
        \includegraphics[clip=true,trim=0mm 0mm 0mm 0mm,width=0.85\linewidth]{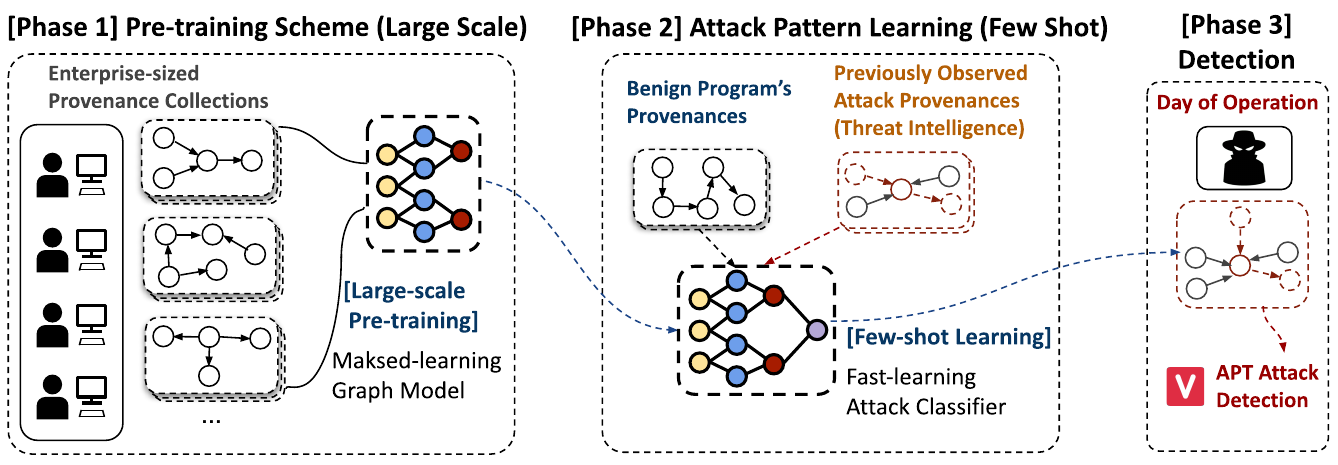}
    }%
    \caption{The workflow of \dupin.}
    \label{fig:workflow}
\end{figure*}

\section{Motivation}\label{sec:motivation}

Attack forensics demands efficient and effective tools to reduce human efforts. In response, several papers have proposed learning-based methods to automatically detect and/or recover APT attack stories. In this context, we use APT attack examples from  \textit{Darpa Transparent Computing (TC)}~\cite{darpa2014trust} as a motivation.

\noindent\textbf{Examples.}
 Consider a scenario where a security analyst uses a learning-based alert system. Assume she has collected audit logs and labeled attack entities. Here, we present a representative sample from both attack (i.e., \textit{attack example - training} in  Figure~\ref{fig:motivation-example}) 
 and benign (i.e., \textit{benign example - training}) activities. Then, we introduce a new attack operation (i.e., \textit{attack example - target}) to illustrate how existing learning systems would work and highlight the need for \dupin{}.
 Provenance graphs of these activities are denoted in Figure~\ref{fig:motivation-example}.

\smallskip
\noindent[\textit{Attack Example - Target}] 
The attack began with a malformed packet sent to a vulnerable \texttt{Nginx} server, allowing the attacker to remotely execute code and establish a shell connection. After gaining access, the attacker 
downloaded an executable file {\tt tmux-1002} and a large number of auxiliary files.
Finally, the attacker initiated an \textit{SSH} channel to acquire a persistence and executed the previously downloaded binary.

\smallskip
\noindent[\textit{Attack Example - Training}] The attack was initiated by a malicious ad server, leading to exploitation. After gaining remote control, the attacker manipulated \texttt{Nginx} to download backdoor files. The attacker then attempted a privilege escalation, which was successful. Finally, the attacker switched to a different shell session as a root user.

\smallskip
\noindent[\textit{Benign Example - Training}] A user visited \textit{Google Drive} and downloaded multiple large files, as shown in (c). \textit{Google Drive} automatically compressed these files into several smaller \texttt{tar.gz} files for download. Before merging, multiple files appear on the audit logs.

\smallskip
The analyst cannot review the entire logs to pinpoint ongoing attacks due to the overwhelming volume of benign system activities. In the remainder of this section, assuming she has deployed learning-based alert systems, we will discuss the expected outcomes of each learning approach.

\noindent\textbf{Supervised Learning Approach.} Traditional approach aims to learn the distribution of both attack and benign activities to distinguish them, using statistical regression/classification methods, or deep learning models~\cite{du2017deeplog,han2020unicorn,alsaheel2021atlas}.

Deep learning systems require a 
large training scale. Likewise, attack-learning methods heavily depend on the scale of available data and the accuracy of labels. In this regard, such data-driven approaches face the challenges of dataset issues. In the wild, the majority of observable audit results are benign, where the real attack activities rarely happen. Moreover, the labeling process is notoriously costly and error-prone, incurring extreme manual efforts. In consequence, they encounter practical constraints due to insufficient data in terms of both quantity and quality. 
As a result, data-driven models may latch onto peripheral, non-causal features of the provenance graph (e.g., specific file names that happened to appear in training attacks) rather than the underlying attack semantics.
For example, the appearance of particular files (e.g., \texttt{`master'}) in the previous attack rather than the critical context can be extracted as a major feature by the model. Consequently, these 
underfit models will incur false positives, resulting in additional human efforts for verification.

\noindent\textbf{Self-supervised Learning Approach.}
After all, the state-of-the-art work methods use self-supervised learning where the model resorts to one class learning (i.e., benign set) and detects attacks as outliers~\cite{ding2023airtag, jia2024magic}. The main idea of this work is to avoid the learning supervision (i.e., attack training) and thereby, remedy the dataset imbalance problem. Instead, it aims to profile the only benign activities and eliminate the supervision in training. The primary issue lies in the complexity in daily use of modern computer systems. The increase in cyber capabilities introduces ambiguity in the benign feature space. Below, we show a real-world case of an app that describes the limitations of self-supervised learning approach by dynamics in benign activities.

Suppose the previous benign activity (\textit{Benign Example - Training}) was added to the training set. Due to the large size, \textit{Google Drive} compresses the files into multiple intermediate parts before transferring. The compression occurs on the server-side, so it is not exhibited to the end user. As a result, the unsupervised models will observe the downloading (i.e., write) of multiple indiscriminately-named files as a benign case. Assume the model then encounters the attack case (\textit{Attack Example - Target}) during the inference stage. In the process of attaining escalated privilege, attacker has downloaded multiple backdoor files. Since this pattern resembles the previous benign activity, the model fails to classify it as abnormal. This case arises because a certain part of the activities of attack (i.e., deploying multiple files) resembles the benign activity. Consequently, the model fails to focus on actual attack context, in this case, creation of a remote shell upon a deployed backdoor.

The ramifications also include false positives. As software evolves rapidly (e.g., through updates or new software), the forensic view of consistent user activities can significantly vary. Thus, one-class learners can struggle with unseen benign patterns, resulting in misclassifications. 

\noindent\textbf{Our Approach.} In recent NLP tasks, the key to fast few-shot training lies in the ability of pre-trained models to understand context. Well-trained language models, such as GPT or BERT-based transformer models require only a small amount of fine-tuning data. Motivated by such success(es) of large language models powered by massive pre-training, we propose an attack classifier model that leverages few shot learning. This method is well-suited for the attack forensics task, considering the extensive log corpus available at the enterprise level for pre-training. A well-pretrained model can understand the behavioral context, allowing downstream task, i.e., few-shot learning on a relatively much smaller amount of data, to quickly converge. To the best of our knowledge, our approach is the first to apply the workflow of (i) massive self-supervised pre-training on benign audit logs and (ii) few shot learning on attack audit logs as a supervised fine-tuning approach to cyber forensics. Figure~\ref{fig:workflow} depicts our workflow.

We revisit the previous example to illustrate why attack learning is needed. Suppose \textit{Attack Training Example} was included in model fine-tuning.
As writing to multiple files appear in both attack and benign cases, the model can unlearn this pattern (as an attack feature). The real attack context (which appears in both attack examples) can hence be abstracted and learned (by fine-tuning) as (\textit{external IP}) $\rightarrow$ (\textit{benign process}) $\rightarrow$ (\textit{randomly named file}) appearing together with (\textit{external IP}) $\rightarrow$ (\textit{remotely accessible process}) $\rightarrow$ (\textit{randomly named file}). Therefore, the original attack could be detected. 
\section{System Design}

\dupin{} primarily comprises four components: (i) tokenizer, (ii) graph constructor, (iii) base graph model and (iv) attack classifier. In this section, we explain the details in each component.

\subsection{Tokenizer}

\noindent\textbf{Token Extraction.} 
A naive tokenization method is to treat audit logs as texts and use existing tokenizers. 
However, it does not capture domain-specific semantics of audit tokens such as {\tt syslog} and {\tt dmesg}. 
In particular, file paths are prevalent in audit logs and many of them have special meanings (e.g., `{\tt /etc/passwd}').
One design choice is to use static method, where the underlying file systems are traversed to collect existing names. While this method ensures comprehensive coverage of tokens which initially present in the file systems, it does not consider token priority and is difficult to evolve. On one hand, it leads to dictionary size overflow, collecting numerous trivial names (e.g., ephemeral, cache files) that are unlikely to reappear.
On the other hand, it can hardly adapt to file system changes.

To mitigate such challenges, we propose a lexical algorithm for dynamic token extraction which runs on a large log corpus. Although this may address the completeness issues, the resulted dictionary may be too large. Therefore, the key lies in reducing the dictionary size by merging redundant names, while preserving significant token names. Assume that we observe two different file paths: `\texttt{/usr/bin/firefox}' and `\texttt{.../cache/.../2A38…54E}' (and many similar files under the same directory). Clearly, `\texttt{firefox}' is a common software that its name should be preserved. In contrast, the latter is one single cache file among many of its type which collectively would generate numerous, yet unnecessary, tokens. Hence, it is essential to merge them and assign a single representative token. We observe that these cache files are repeatedly created and deleted, resulting in appearance of multiple files of the same format within the same directory. Therefore, such mergeable names can be identified by runtime observation. A static snapshot of this directory would reveal a small number of active files at the moment.

In practice, the number of tokens generated by typical logging systems is often very large, primarily due to the redundant appearance of similar names (e.g., cache files). Therefore, we apply a merging strategy until the number of remaining tokens is reduced to the desired size. To determine the file names that ought to be merged and normalized to some special tokens called {\em format tokens}, we first normalize all file names as follows. We replace all characters in file names belonging to the class `\texttt{[a-z|A-Z|0-9]}' with a placeholder `{\tt z}'. For instance, `\texttt{2A38-FFEB}' would be transformed into `\texttt{zzzz-zzzz}'. If a normalized file name appears more than $\kappa$ times in the same directory, meaning that many unique files follow the same format, we consider it a format token and include it in the token dictionary. As such, all files comply with this format are tokenized to the format token. 
Then, we find the appropriate value of $\kappa$, by continuously altering and adding more {\em format tokens} before the size of vocabulary is bounded by $\tau$ (note that the size of vocabulary decreases as $\kappa$ gets higher). The dictionary is then finalized, consisting of a set of format tokens alongside the remaining name tokens. Our tokenizer algorithm ensures coverage of all observed names, either in the form of format tokens or by preserving their original names. However, we use a special token, namely `\texttt{[unknown]}', to address unseen names on-the-fly. We set $\tau$ = 20K for {\tt Linux} logs, 30K for {\tt Windows} logs.

\begin{figure}[H]
    \centering
    {%
    \includegraphics[clip=true,trim=0mm 0mm 0mm 0mm,width=0.8\linewidth]{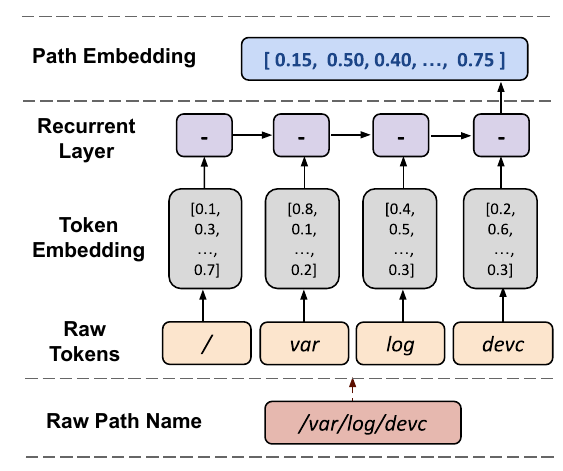}
    }%
	\caption{The file path tokenizer in \dupin.}
	\label{fig:tokenizer}
\end{figure}

\smallskip
\noindent\textbf{Path Embedding.} 
Path embedding refers to the final representation of a file path, created by aggregating the embeddings of its individual tokens. Suppose we have a token embedding; $\mathbf{E}_t: \mathcal{V} \rightarrow \mathbb{R}^F$, where $\mathcal{V}$ is the token dictionary and $F$ is embedding size. First, the file path is split into distinct tokens by each directory, file name or extension (\texttt{Windows} logs). Let these tokens be denoted as $t_1, t_2, ..., t_n$. Then, we use recurrent networks to aggregate these embeddings as $\mathbf{E}_p = \texttt{aggregate}[\{\mathbf{E}_t (t_1), \mathbf{E}_t (t_2), ..., \mathbf{E}_t (t_n)\}]$ (see Figure~\ref{fig:tokenizer}).

The key advantage of recurrent aggregation lies in its ability to consider the sequential nature of file paths. By leveraging the recurrent networks, a path embedding can reflect not only the individual components but also their order and context within the overall path structure. Hence, we use one of the recurrent models, LSTM layer, as the \texttt{aggregate} function. Given that the token sequence in a file path is short, the LSTM appears to be a practical design choice than imposing heavier sequential models. 

\subsection{Graph Construction and Subgraphs Extraction}\label{sec:design-graph-construction}
The goal for training the \dupin{} base model is to allow the model to understand provenance, which is 
a relational representation of system entities (e.g., processes and files), where their connections are based on interactions (e.g., read and execute).
We adhere to a common practice in constructing provenance graphs\cite{yang2023prographer,jia2024magic,rehman2024flash}. On construction, each event leads to the insertion of a corresponding edge, following by Table~\ref{table:edge-type}. We use heterogeneous graphs, where corresponding names (i.e., our path embeddings), edge types are both annotated.  
\begin{table}[h]
\small
\centering
\begin{tabular}{|c|c|}
\hline
\textbf{Edge Types}                                                                                                                    & \textbf{Edge Direction}           \\ \hline \hline
\textit{\begin{tabular}[c]{@{}c@{}} Read, Check File Attribute, \\ Accept, Receive, Load Library\end{tabular}}                          & \texttt{object} $\rightarrow$ \texttt{process} \\ \hline
\textit{\begin{tabular}[c]{@{}c@{}} Send, Write, Bind, Mmap, \\ Connect, Rename, Unlink, \\ Execute, Create, Modify\end{tabular}} & \texttt{process} $\rightarrow$ \texttt{object} \\ \hline
\end{tabular}
\caption{Edge types and edge directions.}
\label{table:edge-type}
\end{table}

Note that we further split the provenance graph into a set of bounded-size subgraphs that collectively cover the full provenance graph. Each subgraph then serves as one training/inference unit for \dupin.
To cultivate the graphs (i.e., inputs to \dupin) from raw logs, we employ a subgraph discovery approach inspired by recent advancements in graph neural networks~\cite{zhou2020graph}.\dupin first iterates the entire set of nodes, ensuring the full coverage. Then, we employ a subgraph search algorithm in which a new neighboring node is randomly added at each step of the exploration process. \dupin employs a threshold-based mechanism to finish the local search, collecting the subgraph once it reaches a predefined size. Additionally ($\gamma$), it sets a separate threshold on the number of iterations to prevent infinite loops during the search process ($\delta$). We set $\gamma = 40$, $\delta = 100$ for this paper. Appendix~\ref{subsec:subgraph-discovery} introduces our subgraph discovery algorithm.

\subsection{Architecture of \dupin{}}

\begin{figure*}[]
    \centering
    {
    \includegraphics[clip=true,trim=0mm 0mm 0mm 0mm,width=0.85\linewidth]{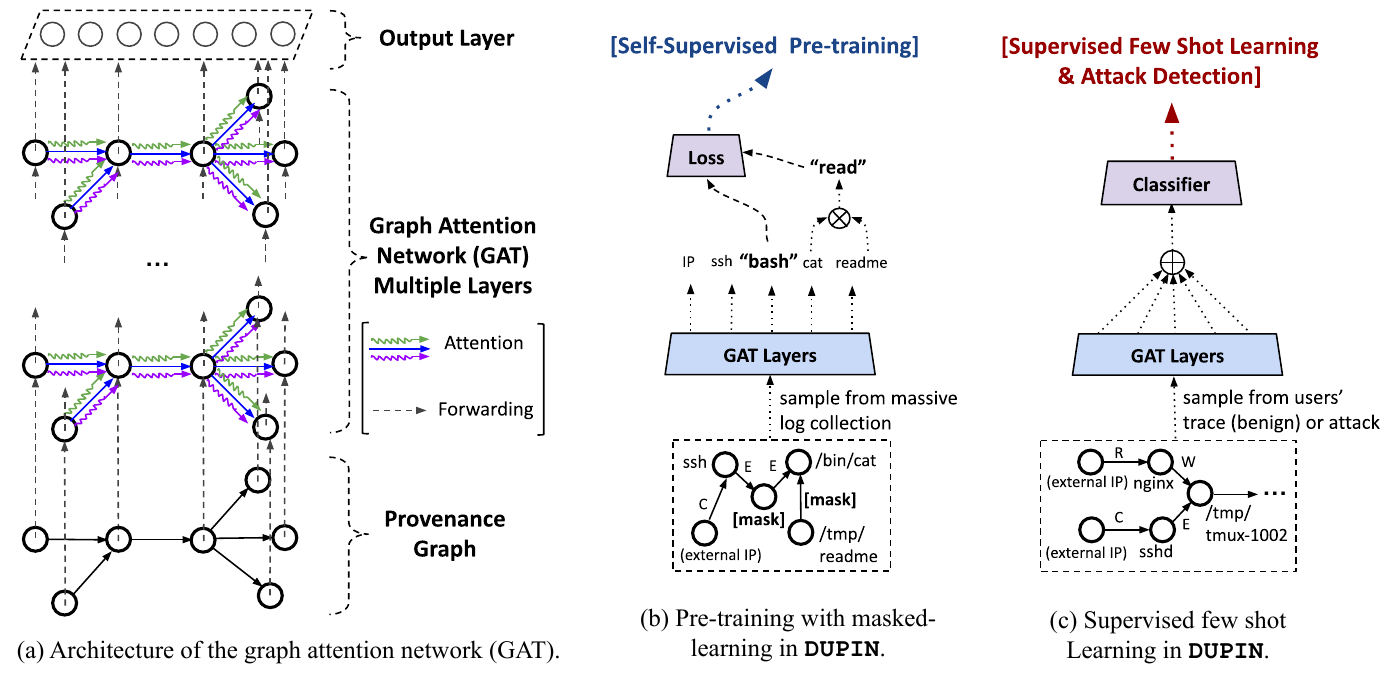}
    }
    \vspace{-1mm}
	\caption{\centering The architecture of \dupin. \dupin{} uses graph attention network (GAT) model as a base (left). For the pre-training, we use graph masked learning (middle). Then, the classifier is attached and trained for the downstream tasks (right).}
	\label{fig:gat}
\end{figure*}

\noindent\textbf{Base Model.} Recent advances in NLP techniques are highly attributed to the attention mechanism~\cite{vaswani2017attention}.
Similarly, we leverage the attention mechanism to enhance interpretability and facilitate effective context understanding. We utilize graph attention networks (GAT) as our base, which leverage the attention mechanism to effectively accommodate the structure of graphs. As shown in Figure~\ref{fig:gat} (left), GAT is a stacked neural structure in which each layer comprises cells representing nodes. At each layer, an attention mechanism is applied along the edges, allowing each node to be influenced by its neighbors according to their importance.

We quote Equations~\eqref{eq:message-forwarding},~\eqref{eq:attention} to illustrate the details and adaptation to provenance graphs~\cite{velivckovic2017graph}. 
A vector $\vec{h}_i'$ for 
a single node $i$ is aggregated from its incoming neighbors (denoted as $\mathcal{N}_i$), weighted by their respective attention.

\begin{equation}
\vec{h}_i' = \sigma \left( \sum_{j \in \mathcal{N}_i} \alpha_{ij} \mathbf{W} \vec{h}_j \right) 
\label{eq:message-forwarding}
\end{equation}

Here, $W \in R^{F \times F}$ is a parameterized weight matrix and $F$ is the cell size of nodes. Importantly, an attention score $\alpha_{ij}$  is applied to each forwarding which is computed as:

\begin{equation}
\alpha_{ij} = \frac{\exp\left( \mathbf{a} \times \left( \mathbf{W} \vec{h}_i \parallel \mathbf{W} \vec{h}_j \right) \right)}{\sum_{k \in \mathcal{N}_i} \exp\left( \mathbf{a} \times \left( \mathbf{W} \vec{h}_i \parallel \mathbf{W} \vec{h}_k \right) \right)}
\label{eq:attention}
\end{equation}

where $\mathbf{a}$ is a shared attention coefficient, $a: R^F \times R^F \rightarrow R$. The attention score between nodes $i$ and $j$ is computed in relation to other neighbor pairs, calculating the comparative score among neighbors. The first layer takes the values of path embeddings, thereby allowing the model to reflect the entity information. However, 
existing models do not utilize edge types (i.e., event names). Therefore, we modify Equation~\eqref{eq:message-forwarding}, adding edge type embeddings, denoted as $z_{ij}$:

\begin{equation}
\vec{h}_i' = \sigma \left( \sum_{j \in \mathcal{N}_i} \alpha_{ij} \left( \mathbf{W} \vec{h}_j + \vec{z}_{ij} \right) \right)
\label{eq:modification}
\end{equation}

We define event embedding $\mathbf{E}_e: E \rightarrow \mathbb{R}^F$, where $E$ is the set of event types (Table~\ref{table:edge-type}). The edge type embedding $z_{ij}$ from $i$ to $j$ is the average of all the event embeddings between $i$ and $j$. Note that there may be multiple events connecting $i$ and $j$.
Specifically, we calculate: 

\begin{equation}
z_{ij} = \frac{1}{N}\sum_{\forall e \in (i, j)} E_e(e)
\end{equation}
where $N$ is the number of different edges from $i$ to $j$. 

In this architecture, \dupin{} harnesses all the information (i.e., entities and events) within provenance graphs, enabling both node- and edge-level heterogeneity. Also, the model can determine the relative importance of entities by utilizing the graph attention mechanisms.

\smallskip
\noindent\textbf{Pretraining.} 
The goal of pretraining is to leverage self-supervised learning to facilitate the model’s ability to understand contexts. For example, BERT-based models~\cite{devlin2018bert} employ the masked language model (MLM) and next sentence prediction (NSP) tasks, which require the model to learn and understand contextual information. We choose node predictions (i.e., process and file names) and edge classification (i.e., event types) as pretraining tasks. Figure~\ref{fig:gat} (middle) shows the example. Here, one can observe the execution flow from the \texttt{SSH} tool to \texttt{cat}, which leads to the speculation that a remote user is attempting to get file content. Once knowing such context, the name `\texttt{bash}' and event `\texttt{read}' can be naturally predicted for the masked places.

Equations~\eqref{eq:node-prediction} and~\eqref{eq:edge-classification} represent the implementation of \dupin{}'s pretraining. Assume $\mathcal{G}_{i}$ and $\mathcal{G}_{ij}$ denote the golds (i.e., answer indexes) of masked node and edge, respectively. We define two parameterized weight matrixes: $W_N \in R^{|V| \times F}$, $W_E \in R^{|E| \times 2F}$ to convert the corresponding output values to one-hot predictions. After applying the \textit{negative log loss} (NLL) and projecting onto the gold labels, we obtain the loss functions.

\begin{equation}
\mathcal{L}_i = - \texttt{log} \left( \texttt{softmax} \left( W_N \times \vec{h_i} \right) \right) \; \vline_{\mathcal{G}_{i}}
\label{eq:node-prediction}
\end{equation}

\begin{equation}
\mathcal{L}_{ij} = - \texttt{log} \left( \texttt{softmax} \left( W_E \times ( \vec{h_i} \mathbin\Vert \vec{h_j} ) \right) \right) \; \vline_{\mathcal{G}_{ij}}
\label{eq:edge-classification}
\end{equation}

Finally, \dupin{} randomly selects masks from the provenance graphs. If multiple masks are selected within a single batch, we sum all losses. 

\begin{equation}
\mathcal{L} = \sum_{i \in {M}_{N}} \mathcal{L}_i + \sum_{(i,j) \in {M}_{E}} L_{ij}
\end{equation}

\subsection{Attack Classifier}
After pre-training, we further fine-tune \dupin{} as an attack classifier. Since the final layer of the pretrained model produces richly contextualized embeddings, we simply add a linear classifier. Given a classification set $C = \{$`\texttt{benign}', `\texttt{attack}'$\}$, we produce a probability output:

\begin{equation}
P_C = \texttt{argmax} \left( \texttt{softmax} \left( \mathbf{W}_{C} \times \sum_{j \in N} \vec{e}_j \right) \right)
\label{eq:classifier}
\end{equation}

In Equation~\eqref{eq:classifier}, $\vec{e}_j$ is the cell values from the final layer and $W_C \in R^{|C| \times F}$ is a weight matrix of the classifier. When training, we aim to minimize alterations to the pretrained cells, preserving the model’s ability to understand context. Therefore, we reduce the learning rate for these cells (Section~\ref{sec:experiment-setup}) to freeze the base model, allowing instead the classifier to adapt to the downstream task. We use \textit{`few-shot learning'} and \textit{`fine-tuning'} interchangeably throughout the paper to refer to this stage.

\subsection{Threat Model}
We assume that the AI component is secure and do not account for the presence of adversarial attacks against our models or the possibility of poisoning during any training phase. \dupin{} is considered part of the trusted computing base.  
Similar to the literature~\cite{du2017deeplog,han2020unicorn,han2021sigl,alsaheel2021atlas,zengy2022shadewatcher,ding2023airtag,yang2023prographer,jia2024magic},
we assume the integrity of audit logging components and log data, even though attackers may engage in long-term operations, resulting in a substantial volume of logs.
\section{Evaluation}

\begin{table*}[]
{
\resizebox{2.05\columnwidth}{!}{%
\begin{tabular}{|c|c|c|c|cccc|cc|}
\hline
\multirow{2}{*}{\textbf{Source}} &
  \multirow{2}{*}{\textbf{Platform}} &
  \multirow{2}{*}{\textbf{ID}} &
  \multicolumn{1}{c|}{\multirow{2}{*}{\textbf{Ref.}}} &
  \multicolumn{4}{c|}{\textbf{Attack Set}} &
  \multicolumn{2}{c|}{\textbf{Benign Set}} \\ \cline{5-10} 
 &
   &
   &
  \multicolumn{1}{l|}{} &
  \multicolumn{1}{c|}{\textbf{Capability}} &
  \multicolumn{1}{c|}{\textbf{Surface}} &
  \multicolumn{1}{c|}{\textbf{Trace Time}} &
  \textbf{\# Entity} &
  \multicolumn{1}{c|}{\textbf{Trace Time}} &
  \textbf{\# Entity} \\ \hline
\multirow{12}{*}{\makecell[c]{\texttt{Darpa}\\ \texttt{Trace}}} &
  \multirow{7}{*}{Linux} &
  1 &
  §3.1 (E3) &
  \multicolumn{1}{c|}{Backdoor} &
  \multicolumn{1}{c|}{\tt nginx} &
  \multicolumn{1}{c|}{04/12, 11:20:00 $\sim$ 11:40:00} &
  9 &
  \multicolumn{1}{c|}{04/03, 11:20:00 $\sim$ 11:40:00} &
  225 \\ \cline{3-10} 
 &
   &
  2 &
  §3.13 (E3) &
  \multicolumn{1}{c|}{Backdoor} &
  \multicolumn{1}{c|}{\tt firefox} &
  \multicolumn{1}{c|}{04/12, 14:00:00 $\sim$ 14:40:00} &
  6 &
  \multicolumn{1}{c|}{04/13, 14:00:00 $\sim$ 14:40:00} &
  423 \\ \cline{3-10} 
 &
   &
  3 &
  §3.14 (E3) &
  \multicolumn{1}{c|}{Backdoor} &
  \multicolumn{1}{c|}{\tt sshd} &
  \multicolumn{1}{c|}{04/13, 09:10:00 $\sim$ 09:20:00} &
  1 &
  \multicolumn{1}{c|}{04/12, 09:10:00 $\sim$ 09:20:00} &
  192 \\ \cline{3-10} 
 &
   &
  4 &
  §3.15 (E3) &
  \multicolumn{1}{c|}{Dropper} &
  \multicolumn{1}{c|}{\tt pine} &
  \multicolumn{1}{c|}{04/13, 12:45:00 $\sim$ 12:55:00} &
  2 &
  \multicolumn{1}{c|}{04/13, 10:45:00 $\sim$ 10:55:00} &
  2,756 \\ \cline{3-10} 
 &
   &
  5 &
  §4.9 (E3) &
  \multicolumn{1}{c|}{Phishing} &
  \multicolumn{1}{c|}{\tt pine} &
  \multicolumn{1}{c|}{04/13, 14:00:00 $\sim$ 14:30:00} &
  3 &
  \multicolumn{1}{c|}{04/13, 10:00:00 $\sim$ 10:30:00} &
  3,285 \\ \cline{3-10} 
 &
   &
  6 &
  §7.3 (E5) &
  \multicolumn{1}{c|}{Backdoor} &
  \multicolumn{1}{c|}{\tt firefox} &
  \multicolumn{1}{c|}{05/13, 11:00:00 $\sim$ 11:10:00} &
  4 &
  \multicolumn{1}{c|}{05/07, 11:00:00 $\sim$ 11:10:00} &
  198 \\ \cline{3-10} 
 &
   &
  7 &
  §10.4 (E5) &
  \multicolumn{1}{c|}{Exfiltration} &
  \multicolumn{1}{c|}{\tt nginx} &
  \multicolumn{1}{c|}{05/09, 15:25:00 $\sim$ 16:05:00} &
  7 &
  \multicolumn{1}{c|}{05/07, 15:25:00 $\sim$ 16:05:00} &
  134 \\ \cline{2-10} 
 &
  \multirow{5}{*}{Windows} &
  8 &
  §3.4 (E3) &
  \multicolumn{1}{c|}{Backdoor} &
  \multicolumn{1}{c|}{\tt firefox} &
  \multicolumn{1}{c|}{04/11, 10:08:42 $\sim$ 10:38:42} &
  13 &
  \multicolumn{1}{c|}{04/05, 10:08:42 $\sim$ 10:38:42} &
  251 \\ \cline{3-10} 
 &
   &
  9 &
  §3.10 (E3) &
  \multicolumn{1}{c|}{Dropper} &
  \multicolumn{1}{c|}{\tt firefox} &
  \multicolumn{1}{c|}{04/12, 11:10:00 $\sim$ 11:20:00} &
  4 &
  \multicolumn{1}{c|}{04/05, 11:10:00 $\sim$ 11:20:00} &
  118 \\ \cline{3-10} 
 &
   &
  10 &
  §4.4 (E3) &
  \multicolumn{1}{c|}{Phishing} &
  \multicolumn{1}{c|}{\tt powershell} &
  \multicolumn{1}{c|}{04/09, 14:50:00 $\sim$ 15:10:00} &
  50 &
  \multicolumn{1}{c|}{04/04, 14:50:00 $\sim$ 15:10:00} &
  78 \\ \cline{3-10} 
 &
   &
  11 &
  §9.4 (E5) &
  \multicolumn{1}{c|}{Backdoor} &
  \multicolumn{1}{c|}{\tt firefox} &
  \multicolumn{1}{c|}{05/16, 15:00:00 $\sim$ 15:20:00} &
  6 &
  \multicolumn{1}{c|}{05/07, 15:00:00 $\sim$ 15:20:00} &
  402 \\ \cline{3-10} 
 &
   &
  12 &
  §10.6 (E5) &
  \multicolumn{1}{c|}{Elevate} &
  \multicolumn{1}{c|}{\tt firefox} &
  \multicolumn{1}{c|}{05/17, 12:05:00 $\sim$ 12:30:00} &
  11 &
  \multicolumn{1}{c|}{05/07, 12:05:00 $\sim$ 12:30:00} &
  87 \\ \hline
\multirow{6}{*}{\makecell[c]{\texttt{ATLAS}}} &
  \multirow{6}{*}{Windows} &
  13 &
  M-1 &
  \multicolumn{1}{c|}{Compromise} &
  \multicolumn{1}{c|}{-} &
  \multicolumn{1}{c|}{12/21, 16:27:07 $\sim$ 16:51:10} &
  2 &
  \multicolumn{1}{c|}{11/02 22:28:09 $\sim$ 22:47:15} &
  33 \\ \cline{3-10} 
 &
   &
  14 &
  M-2 &
  \multicolumn{1}{c|}{Phishing} &
  \multicolumn{1}{c|}{-} &
  \multicolumn{1}{c|}{12/22, 09:10:00 $\sim$ 09:30:00} &
  1 &
  \multicolumn{1}{c|}{11/02 22:28:09 $\sim$ 22:47:15} &
  33 \\ \cline{3-10} 
 &
   &
  15 &
  M-3 &
  \multicolumn{1}{c|}{Maladvertising} &
  \multicolumn{1}{c|}{-} &
  \multicolumn{1}{c|}{12/21, 18:45:00 $\sim$ 18:55:00} &
  6 &
  \multicolumn{1}{c|}{11/27 11:49:31 $\sim$ 11:53:07} &
  31 \\ \cline{3-10} 
 &
   &
  16 &
  M-4 &
  \multicolumn{1}{c|}{Monero Miner} &
  \multicolumn{1}{c|}{-} &
  \multicolumn{1}{c|}{12/22, 00:00:00 $\sim$ 00:10:00} &
  9 &
  \multicolumn{1}{c|}{12/01 14:19:48 $\sim$ 14:38:27} &
  46 \\ \cline{3-10} 
 &
   &
  17 &
  M-5 &
  \multicolumn{1}{c|}{Compromise} &
  \multicolumn{1}{c|}{-} &
  \multicolumn{1}{c|}{12/15, 15:10:00 $\sim$ 15:20:00} &
  11 &
  \multicolumn{1}{c|}{12/15 10:51:11 $\sim$ 15:20:00} &
  68 \\ \cline{3-10} 
 &
   &
  18 &
  M-6 &
  \multicolumn{1}{c|}{Spam} &
  \multicolumn{1}{c|}{-} &
  \multicolumn{1}{c|}{12/21, 20:20:00 $\sim$ 20:30:10} &
  31 &
  \multicolumn{1}{c|}{12/01 14:19:49 $\sim$ 14:38:00} &
  41 \\ \hline
\multirow{7}{*}{\makecell[c]{\texttt{Palantir}}} &
  \multirow{7}{*}{Linux} &
  19 &
  \multirow{2}{*}{§2.1} &
  \multicolumn{1}{c|}{\multirow{2}{*}{Watering Hole}} &
  \multicolumn{1}{c|}{\tt wget} &
  \multicolumn{1}{c|}{\multirow{2}{*}{04/19, 12:38:40 $\sim$ 12:44:52}} &
  15 &
  \multicolumn{1}{c|}{\multirow{2}{*}{04/19, 12:38:40 $\sim$ 12:44:52}} &
  \multirow{2}{*}{1,010} \\ \cline{3-3} \cline{6-6} \cline{8-8}
 &
   &
  20 &
   &
  \multicolumn{1}{c|}{} &
  \multicolumn{1}{c|}{\tt nginx} &
  \multicolumn{1}{c|}{} &
  15 &
  \multicolumn{1}{c|}{} &
   \\ \cline{3-10} 
 &
   &
  21 &
  \multirow{2}{*}{§8.1.1} &
  \multicolumn{1}{c|}{\multirow{2}{*}{Data Leakage}} &
  \multicolumn{1}{c|}{curl} &
  \multicolumn{1}{c|}{\multirow{2}{*}{04/19, 13:04:59 $\sim$ 13:09:38}} &
  107 &
  \multicolumn{1}{c|}{\multirow{2}{*}{04/19, 13:04:59 $\sim$ 13:09:38}} &
  \multirow{2}{*}{609} \\ \cline{3-3} \cline{6-6} \cline{8-8}
 &
   &
  22 &
   &
  \multicolumn{1}{c|}{} &
  \multicolumn{1}{c|}{\tt ftpd} &
  \multicolumn{1}{c|}{} &
  78 &
  \multicolumn{1}{c|}{} &
   \\ \cline{3-10} 
 &
   &
  23 &
  \multirow{2}{*}{§8.1.2} &
  \multicolumn{1}{c|}{\multirow{2}{*}{Insider Threat}} &
  \multicolumn{1}{c|}{\tt cp} &
  \multicolumn{1}{c|}{\multirow{2}{*}{04/19, 12:59:03 $\sim$ 13:03:02}} &
  510 &
  \multicolumn{1}{c|}{\multirow{2}{*}{04/19, 12:59:03 $\sim$ 13:03:02}} &
  \multirow{2}{*}{585} \\ \cline{3-3} \cline{6-6} \cline{8-8}
 &
   &
  24 &
   &
  \multicolumn{1}{c|}{} &
  \multicolumn{1}{c|}{\tt lighttpd} &
  \multicolumn{1}{c|}{} &
  163 &
  \multicolumn{1}{c|}{} &
   \\ \cline{3-10} 
 &
   &
  25 &
  §A.D &
  \multicolumn{1}{c|}{Phishing} &
  \multicolumn{1}{c|}{\tt sendmail} &
  \multicolumn{1}{c|}{04/20, 04:34:40 $\sim$ 04:37:30} &
  13 &
  \multicolumn{1}{c|}{04/20, 04:34:40 $\sim$ 04:37:30} &
  262 \\ \hline
\end{tabular}%
}
}
\caption{Summary of the dataset.}
\label{table:dataset-summary}
\end{table*}

We aim to ensure that our experiment accurately reflects real-world conditions in attack surveillance. Since subtle experimental pitfalls can lead to misaligned evaluation results~\cite{arp2022and,liu2025we}, we carefully design and adhere to the following experimental principles. 

First, it is essential that a model focuses on fundamental attack contexts and avoids learning or relying on non-fundamental, peripheral cues that may coexist across both the training and validation datasets. As attack datasets contain multiple distinct attack attempts, they are typically partitioned into cross-validation sets for model training and testing. However, although these sets are assumed to be distinct, they are often generated by the same red team and on a fixed victim environment, resulting in peripheral but non-trivial similarities across the validation splits, such as recurring system-wide artifacts from shared machine configurations or unintended human-level fingerprints. These subtle consistencies can ultimately lead to significant data leakage, compromising the validity of the evaluation. To address this issue, we propose that datasets should be split for training and testing based on attack authorship (i.e., by distinct dataset sources). We adopt this principle in our experiments as it better reflects realistic deployment scenarios, where observable attacks continuously evolve, hence mutate and are carried out by diverse actors.

Second, the entire workflow prior to training and/or prediction must remain entirely independent of ground-truth information. A standard forensic process entails significant amount of data processing, converting raw textual logs into structured graphs annotated with named tokens, which may introduce ambiguity during the labeling phase. Unfortunately, there is no commonly standardized workflow for collecting finalized data prior to model inference, despite its impact on model performance in some degree. For instance, one may mistakenly allow search algorithm to start from an attack entity or set the inclusion of attack entities as a condition for search termination. Here, our second principle is that the attack ground truth must not be utilized in any form. We unify the search algorithm, implementation, and parameters for collecting both attack and benign samples without use of ground-truth information in any form.

Third, the data distribution needs to mirror the real-world characteristics. Existing attack datasets naturally inherit the problem of class imbalance, where the model suffers from limited attack samples. Rather than addressing this imbalance through artificial sampling techniques, we choose to embrace the intrinsic nature of attack surveillance. Hence, our training and validation data remain intentionally unaltered. 

Last, it is important to employ proper evaluation metrics. For instance, due to the class imbalance problem, accuracy rate can significantly overestimate model performance by significantly ignoring false negatives. In contrast, classification tasks in the security domain typically favor precision and recall, as these metrics equally capture a degree of both attack fatigue and false silence. However, under severe class imbalance, even a small proportion of false positives can dominate the overall evaluation, potentially blinding the improvements in true positive rate (i.e., attack detection). If using the classification accuracy, false negatives can be unnoticed due to data imbalance. To address this, we explicitly report both true positives (i.e., true positive rate) and false positives (i.e., true negative rate), so that the whole evaluation metrics in confusion matrix can be recovered. Additionally, we incorporate the receiver operating characteristic (ROC) curve and the area under the ROC curve (AUROC), which better illustrate the trade-off between true positives and false positives.

\subsection{Dataset}
\noindent\textbf{Overview.} We collect the dataset from multiple sources to ensure that the training and testing sets are strictly separated based on attack authorship. In this context, the most reliable definition of an attack author corresponds to the dataset source itself, where training and testing sets originate from entirely different sources. We have verified that each dataset source employs distinct attack references, which helps prevent data leakage and allows us to treat them as independent authors within our experimental design. 

First, we use the DARPA Transparent Computing ({\tt DARPA TRACE}) program~\cite{darpa2014trust}, which is widely adopted in the intrusion detection literature~\cite{jia2024magic,rehman2024flash,cheng2024kairos,bilot2025sometimes,zhang2025tapas}.
We use two sub-versions, namely E3 and E5, which we treat as originating from a single data source. The {\tt DARPA TRACE} dataset includes tracepoints collected from both {\tt Windows} and {\tt Linux} platforms, including 12 attack sets. 
We include \texttt{Windows} and \texttt{Linux} tracepoints (i.e., {\tt Trace}, {\tt Cadets} and {\tt Fivedirections}). \texttt{Darpa TRACE} dataset includes one additional collection from the mobile environment setting ({\tt Clearscope}), which we omit due to limitations in collecting extensive pretraining data and name tokens and difficulties in integrating it with other dataset sources.
Additionally, we incorporate two more datasets: a {\tt Windows}-based attack dataset by {\tt ATLAS}~\cite{alsaheel2021atlas} and a {\tt Linux}-based attack dataset from {\tt Palantir}~\cite{zeng2022palantir}. As a result, our evaluation includes two distinct dataset sources for each platform: {\tt Windows} and {\tt Linux}. Table~\ref{table:dataset-summary} presents the complete set of datasets used in our study, comprising 25 distinct APT campaigns collected from four different sources. 
This diversity is intended to support a comprehensive evaluation and prevent models and configurations (e.g., hyperparameters) of \dupin and baselines from overfitting to any single dataset.

\noindent\textbf{Attack Set.} 
The attack set is curated based on the official documentation provided with each dataset release (see Table~\ref{table:dataset-summary} for corresponding references). As a first step, we apply Algorithm~\ref{alg:subgraph-discovery} directly to the raw logs to extract subgraphs, without use of ground-truth annotations, in accordance with our evaluation principle. Afterward, we identify attack entities—those that contain at least one attack events, the edges that represent known malicious activities. In real-world surveillance, attack entities are inherently interwoven with benign activities throughout the system trace. Our attack collecting process reflects such characteristics of nature and attack behaviors are shown partially along with other benign activities.

As a reference to attack sets, we include the section numbers that we reference to identify the attacks within each dataset. We also include the number of entities (i.e., that of subgraphs), the time of attack period of each attack campaign in Table~\ref{table:dataset-summary}.

{\tt DARPA TRACE} dataset provides extensive logs from multiple tracepoints, with attack activities annotated by their operational time windows and the names of related entities, rather the fixed size of time window per each attack set. We carefully construct the attack sets by referencing the official documentation. Details regarding the usage of {\tt DARPA TRACE} dataset are provided in Section~\ref{subsec:usage-darpa-dataset}.

\noindent\textbf{Benign Set.} 
We directly rely on Algorithm~\ref{alg:subgraph-discovery} to construct the benign set. The key is use of a unified algorithm for collecting both attack and benign sets, ensuring consistency and preventing bias. 
Alsaheel et al.~\cite{alsaheel2021atlas} provides time-separated traces for the attack and benign sets for each campaign, which we directly classify into their respective categories. The {\tt DARPA TRACE} dataset provides a continuous log trace, with red team activities annotated according to their operational time windows. To eliminate any possibility of attack contamination in our benign sets, we enforce a strict temporal separation between attack and benign periods. Specifically, we use the first day of the corresponding trace to collect benign data. Additionally, we ensure that the duration and clock time of the benign trace match those of the corresponding attack trace. Finally, \texttt{Palantir}~\cite{zeng2022palantir} provides a single trace covering the entire duration per each attack operation. To isolate benign behavior, we apply a logical separation by removing any benign processes that are used for attack purposes during the attack period. Note that for attack sets of ID 19 -- 24 (see Table~\ref{table:dataset-summary}), each couple of sets shares one single benign set. Also, \texttt{M-5} from Alsaheel et al.~\cite{alsaheel2021atlas} dataset is exceptionally processed in this manner, (i.e., logical separation) as no distinct trace is provided for attack and benign activities.

\begin{table*}[]
{
\centering
\resizebox{2.05\columnwidth}{!}{%
\begin{tabular}{|c|ccccccc|ccccccc|ccccccc|}
\hline
 &
  \multicolumn{7}{c|}{\textbf{\dupin}} &
  \multicolumn{7}{c|}{\textbf{MAGIC}} &
  \multicolumn{7}{c|}{\textbf{AirTag}} \\ \cline{2-22} 
\multirow{-2}{*}{\textbf{ID}} &
  \textbf{TP} &
  \textbf{FN} &
  \textbf{TN} &
  \textbf{FP} &
  \textbf{TPR} &
  \textbf{TNR} &
  \textbf{AUR.} &
  \textbf{TP} &
  \multicolumn{1}{l}{\textbf{FN}} &
  \textbf{TN} &
  \multicolumn{1}{l}{\textbf{FP}} &
  \multicolumn{1}{l}{\textbf{TPR}} &
  \multicolumn{1}{l}{\textbf{TNR}} &
  \textbf{AUR.} &
  \textbf{TP} &
  \textbf{FN} &
  \textbf{TN} &
  \textbf{FP} &
  \textbf{TPR} &
  \textbf{TNR} &
  \textbf{AUR.} \\ \hline
1 &
  9 &
  0 &
  170 &
  55 &
  \textbf{1.00} &
  \textbf{0.76} &
  \textbf{0.97} &
  8 &
  1 &
  134 &
  91 &
  0.89 &
  0.60 &
  0.83 &
  5 &
  4 &
  64 &
  161 &
  0.56 &
  0.28 &
  0.31 \\ \cline{1-1}
2 &
  \cellcolor[HTML]{EFEFEF}5 &
  \cellcolor[HTML]{EFEFEF}1 &
  \cellcolor[HTML]{EFEFEF}345 &
  \cellcolor[HTML]{EFEFEF}78 &
  \cellcolor[HTML]{EFEFEF}0.83 &
  \cellcolor[HTML]{EFEFEF}\textbf{0.82} &
  \cellcolor[HTML]{EFEFEF}0.80 &
  \cellcolor[HTML]{EFEFEF}6 &
  \cellcolor[HTML]{EFEFEF}0 &
  \cellcolor[HTML]{EFEFEF}243 &
  \cellcolor[HTML]{EFEFEF}180 &
  \cellcolor[HTML]{EFEFEF}\textbf{1.00} &
  \cellcolor[HTML]{EFEFEF}0.57 &
  \cellcolor[HTML]{EFEFEF}\textbf{0.94} &
  \cellcolor[HTML]{EFEFEF}1 &
  \cellcolor[HTML]{EFEFEF}5 &
  \cellcolor[HTML]{EFEFEF}91 &
  \cellcolor[HTML]{EFEFEF}332 &
  \cellcolor[HTML]{EFEFEF}0.17 &
  \cellcolor[HTML]{EFEFEF}0.22 &
  \cellcolor[HTML]{EFEFEF}0.08 \\ \cline{1-1}
3 &
  1 &
  0 &
  145 &
  47 &
  \textbf{1.00} &
  \textbf{0.76} &
  \textbf{0.99} &
  1 &
  0 &
  122 &
  70 &
  \textbf{1.00} &
  0.64 &
  0.79 &
  0 &
  1 &
  28 &
  164 &
  0.00 &
  0.15 &
  0.12 \\ \cline{1-1}
4 &
  \cellcolor[HTML]{EFEFEF}2 &
  \cellcolor[HTML]{EFEFEF}0 &
  \cellcolor[HTML]{EFEFEF}2,630 &
  \cellcolor[HTML]{EFEFEF}126 &
  \cellcolor[HTML]{EFEFEF}\textbf{1.00} &
  \cellcolor[HTML]{EFEFEF}\textbf{0.95} &
  \cellcolor[HTML]{EFEFEF}\textbf{0.99} &
  \cellcolor[HTML]{EFEFEF}2 &
  \cellcolor[HTML]{EFEFEF}0 &
  \cellcolor[HTML]{EFEFEF}2,279 &
  \cellcolor[HTML]{EFEFEF}477 &
  \cellcolor[HTML]{EFEFEF}\textbf{1.00} &
  \cellcolor[HTML]{EFEFEF}0.83 &
  \cellcolor[HTML]{EFEFEF}0.93 &
  \cellcolor[HTML]{EFEFEF}2 &
  \cellcolor[HTML]{EFEFEF}0 &
  \cellcolor[HTML]{EFEFEF}2,442 &
  \cellcolor[HTML]{EFEFEF}314 &
  \cellcolor[HTML]{EFEFEF}\textbf{1.00} &
  \cellcolor[HTML]{EFEFEF}0.89 &
  \cellcolor[HTML]{EFEFEF}0.94 \\ \cline{1-1}
5 &
  0 &
  3 &
  3,266 &
  19 &
  0.00 &
  \textbf{0.99} &
  0.65 &
  0 &
  3 &
  {\color[HTML]{333333} 2,731} &
  554 &
  0.00 &
  0.83 &
  \textbf{0.79} &
  0 &
  3 &
  {\color[HTML]{333333} 2,689} &
  {\color[HTML]{333333} 596} &
  0.00 &
  0.82 &
  0.32 \\ \cline{1-1}
6 &
  \cellcolor[HTML]{EFEFEF}0 &
  \cellcolor[HTML]{EFEFEF}4 &
  \cellcolor[HTML]{EFEFEF}197 &
  \cellcolor[HTML]{EFEFEF}54 &
  \cellcolor[HTML]{EFEFEF}0.00 &
  \cellcolor[HTML]{EFEFEF}\textbf{0.99} &
  \cellcolor[HTML]{EFEFEF}0.41 &
  \cellcolor[HTML]{EFEFEF}0 &
  \cellcolor[HTML]{EFEFEF}4 &
  \cellcolor[HTML]{EFEFEF}190 &
  \cellcolor[HTML]{EFEFEF}8 &
  \cellcolor[HTML]{EFEFEF}0.00 &
  \cellcolor[HTML]{EFEFEF}0.96 &
  \cellcolor[HTML]{EFEFEF}\textbf{0.49} &
  \cellcolor[HTML]{EFEFEF}4 &
  \cellcolor[HTML]{EFEFEF}0 &
  \cellcolor[HTML]{EFEFEF}49 &
  \cellcolor[HTML]{EFEFEF}149 &
  \cellcolor[HTML]{EFEFEF}\textbf{1.00} &
  \cellcolor[HTML]{EFEFEF}0.25 &
  \cellcolor[HTML]{EFEFEF}0.25 \\ \cline{1-1}
7 &
  7 &
  0 &
  111 &
  23 &
  \textbf{1.00} &
  \textbf{0.83} &
  \textbf{0.94} &
  5 &
  2 &
  83 &
  51 &
  0.71 &
  0.62 &
  0.66 &
  7 &
  0 &
  49 &
  85 &
  \textbf{1.00} &
  0.37 &
  0.51 \\ \cline{1-1}
8 &
  \cellcolor[HTML]{EFEFEF}11 &
  \cellcolor[HTML]{EFEFEF}2 &
  \cellcolor[HTML]{EFEFEF}197 &
  \cellcolor[HTML]{EFEFEF}54 &
  \cellcolor[HTML]{EFEFEF}0.85 &
  \cellcolor[HTML]{EFEFEF}\textbf{0.78} &
  \cellcolor[HTML]{EFEFEF}\textbf{0.87} &
  \cellcolor[HTML]{EFEFEF}13 &
  \cellcolor[HTML]{EFEFEF}0 &
  \cellcolor[HTML]{EFEFEF}16 &
  \cellcolor[HTML]{EFEFEF}235 &
  \cellcolor[HTML]{EFEFEF}\textbf{1.00} &
  \cellcolor[HTML]{EFEFEF}0.06 &
  \cellcolor[HTML]{EFEFEF}0.17 &
  \cellcolor[HTML]{EFEFEF}6 &
  \cellcolor[HTML]{EFEFEF}7 &
  \cellcolor[HTML]{EFEFEF}157 &
  \cellcolor[HTML]{EFEFEF}94 &
  \cellcolor[HTML]{EFEFEF}0.46 &
  \cellcolor[HTML]{EFEFEF}0.63 &
  \cellcolor[HTML]{EFEFEF}0.35 \\ \cline{1-1}
9 &
  1 &
  3 &
  80 &
  38 &
  0.25 &
  \textbf{0.68} &
  0.45 &
  2 &
  2 &
  12 &
  106 &
  0.50 &
  0.10 &
  0.11 &
  4 &
  0 &
  67 &
  51 &
  \textbf{1.00} &
  0.57 &
  \textbf{0.57} \\ \cline{1-1}
10 &
  \cellcolor[HTML]{EFEFEF}34 &
  \cellcolor[HTML]{EFEFEF}16 &
  \cellcolor[HTML]{EFEFEF}53 &
  \cellcolor[HTML]{EFEFEF}25 &
  \cellcolor[HTML]{EFEFEF}0.68 &
  \cellcolor[HTML]{EFEFEF}\textbf{0.68} &
  \cellcolor[HTML]{EFEFEF}\textbf{0.68} &
  \cellcolor[HTML]{EFEFEF}46 &
  \cellcolor[HTML]{EFEFEF}4 &
  \cellcolor[HTML]{EFEFEF}10 &
  \cellcolor[HTML]{EFEFEF}68 &
  \cellcolor[HTML]{EFEFEF}\textbf{0.92} &
  \cellcolor[HTML]{EFEFEF}0.13 &
  \cellcolor[HTML]{EFEFEF}0.14 &
  \cellcolor[HTML]{EFEFEF}0 &
  \cellcolor[HTML]{EFEFEF}50 &
  \cellcolor[HTML]{EFEFEF}40 &
  \cellcolor[HTML]{EFEFEF}38 &
  \cellcolor[HTML]{EFEFEF}0.00 &
  \cellcolor[HTML]{EFEFEF}0.51 &
  \cellcolor[HTML]{EFEFEF}0.14 \\ \cline{1-1}
11 &
  6 &
  0 &
  356 &
  46 &
  \textbf{1.00} &
  \textbf{0.89} &
  \textbf{0.99} &
  6 &
  0 &
  126 &
  276 &
  \textbf{1.00} &
  0.31 &
  0.64 &
  0 &
  6 &
  255 &
  147 &
  0.00 &
  0.63 &
  0.00 \\ \cline{1-1}
12 &
  \cellcolor[HTML]{EFEFEF}5 &
  \cellcolor[HTML]{EFEFEF}6 &
  \cellcolor[HTML]{EFEFEF}74 &
  \cellcolor[HTML]{EFEFEF}13 &
  \cellcolor[HTML]{EFEFEF}0.45 &
  \cellcolor[HTML]{EFEFEF}\textbf{0.85} &
  \cellcolor[HTML]{EFEFEF}0.70 &
  \cellcolor[HTML]{EFEFEF}11 &
  \cellcolor[HTML]{EFEFEF}0 &
  \cellcolor[HTML]{EFEFEF}16 &
  \cellcolor[HTML]{EFEFEF}71 &
  \cellcolor[HTML]{EFEFEF}\textbf{1.00} &
  \cellcolor[HTML]{EFEFEF}0.18 &
  \cellcolor[HTML]{EFEFEF}0.48 &
  \cellcolor[HTML]{EFEFEF}11 &
  \cellcolor[HTML]{EFEFEF}0 &
  \cellcolor[HTML]{EFEFEF}68 &
  \cellcolor[HTML]{EFEFEF}19 &
  \cellcolor[HTML]{EFEFEF}\textbf{1.00} &
  \cellcolor[HTML]{EFEFEF}0.78 &
  \cellcolor[HTML]{EFEFEF}\textbf{0.84} \\ \cline{1-1}
13 &
  2 &
  0 &
  15 &
  18 &
  \textbf{1.00} &
  0.45 &
  \textbf{0.74} &
  1 &
  1 &
  16 &
  17 &
  0.50 &
  0.48 &
  0.38 &
  0 &
  2 &
  23 &
  10 &
  0.00 &
  0.70 &
  0.36 \\ \cline{1-1}
14 &
  \cellcolor[HTML]{EFEFEF}1 &
  \cellcolor[HTML]{EFEFEF}0 &
  \cellcolor[HTML]{EFEFEF}10 &
  \cellcolor[HTML]{EFEFEF}21 &
  \cellcolor[HTML]{EFEFEF}\textbf{1.00} &
  \cellcolor[HTML]{EFEFEF}0.32 &
  \cellcolor[HTML]{EFEFEF}\textbf{0.84} &
  \cellcolor[HTML]{EFEFEF}0 &
  \cellcolor[HTML]{EFEFEF}1 &
  \cellcolor[HTML]{EFEFEF}14 &
  \cellcolor[HTML]{EFEFEF}17 &
  \cellcolor[HTML]{EFEFEF}0.00 &
  \cellcolor[HTML]{EFEFEF}0.45 &
  \cellcolor[HTML]{EFEFEF}0.23 &
  \cellcolor[HTML]{EFEFEF}0 &
  \cellcolor[HTML]{EFEFEF}1 &
  \cellcolor[HTML]{EFEFEF}23 &
  \cellcolor[HTML]{EFEFEF}8 &
  \cellcolor[HTML]{EFEFEF}0.00 &
  \cellcolor[HTML]{EFEFEF}0.74 &
  \cellcolor[HTML]{EFEFEF}0.48 \\ \cline{1-1}
15 &
  3 &
  3 &
  15 &
  16 &
  0.50 &
  0.48 &
  \textbf{0.62} &
  4 &
  2 &
  13 &
  18 &
  \textbf{0.67} &
  0.42 &
  0.45 &
  2 &
  4 &
  23 &
  8 &
  0.33 &
  0.74 &
  0.60 \\ \cline{1-1}
16 &
  \cellcolor[HTML]{EFEFEF}7 &
  \cellcolor[HTML]{EFEFEF}2 &
  \cellcolor[HTML]{EFEFEF}26 &
  \cellcolor[HTML]{EFEFEF}20 &
  \cellcolor[HTML]{EFEFEF}\textbf{0.78} &
  \cellcolor[HTML]{EFEFEF}0.57 &
  \cellcolor[HTML]{EFEFEF}\textbf{0.69} &
  \cellcolor[HTML]{EFEFEF}5 &
  \cellcolor[HTML]{EFEFEF}4 &
  \cellcolor[HTML]{EFEFEF}21 &
  \cellcolor[HTML]{EFEFEF}25 &
  \cellcolor[HTML]{EFEFEF}0.56 &
  \cellcolor[HTML]{EFEFEF}0.46 &
  \cellcolor[HTML]{EFEFEF}0.50 &
  \cellcolor[HTML]{EFEFEF}0 &
  \cellcolor[HTML]{EFEFEF}9 &
  \cellcolor[HTML]{EFEFEF}31 &
  \cellcolor[HTML]{EFEFEF}15 &
  \cellcolor[HTML]{EFEFEF}0.00 &
  \cellcolor[HTML]{EFEFEF}0.67 &
  \cellcolor[HTML]{EFEFEF}0.27 \\ \cline{1-1}
17 &
  9 &
  2 &
  44 &
  24 &
  \textbf{0.82} &
  \textbf{0.65} &
  \textbf{0.69} &
  7 &
  4 &
  24 &
  44 &
  0.64 &
  0.35 &
  0.41 &
  8 &
  3 &
  36 &
  32 &
  0.73 &
  0.53 &
  0.44 \\ \cline{1-1}
18 &
  \cellcolor[HTML]{EFEFEF}21 &
  \cellcolor[HTML]{EFEFEF}10 &
  \cellcolor[HTML]{EFEFEF}23 &
  \cellcolor[HTML]{EFEFEF}18 &
  \cellcolor[HTML]{EFEFEF}\textbf{0.68} &
  \cellcolor[HTML]{EFEFEF}0.56 &
  \cellcolor[HTML]{EFEFEF}\textbf{0.66} &
  \cellcolor[HTML]{EFEFEF}16 &
  \cellcolor[HTML]{EFEFEF}15 &
  \cellcolor[HTML]{EFEFEF}16 &
  \cellcolor[HTML]{EFEFEF}25 &
  \cellcolor[HTML]{EFEFEF}0.52 &
  \cellcolor[HTML]{EFEFEF}0.39 &
  \cellcolor[HTML]{EFEFEF}0.35 &
  \cellcolor[HTML]{EFEFEF}0 &
  \cellcolor[HTML]{EFEFEF}31 &
  \cellcolor[HTML]{EFEFEF}27 &
  \cellcolor[HTML]{EFEFEF}14 &
  \cellcolor[HTML]{EFEFEF}0.00 &
  \cellcolor[HTML]{EFEFEF}0.66 &
  \cellcolor[HTML]{EFEFEF}0.04 \\ \cline{1-1}
19 &
  8 &
  7 &
   &
   &
  0.53 &
   &
  \textbf{0.66} &
  4 &
  11 &
   &
   &
  0.27 &
   &
  0.35 &
  15 &
  0 &
   &
   &
  \textbf{1.00} &
   &
  0.47 \\ \cline{1-1}
20 &
  \cellcolor[HTML]{EFEFEF}13 &
  \cellcolor[HTML]{EFEFEF}2 &
  \multirow{-2}{*}{607} &
  \multirow{-2}{*}{403} &
  \cellcolor[HTML]{EFEFEF}\textbf{0.87} &
  \multirow{-2}{*}{0.60} &
  \cellcolor[HTML]{EFEFEF}0.77 &
  \cellcolor[HTML]{EFEFEF}11 &
  \cellcolor[HTML]{EFEFEF}4 &
  \multirow{-2}{*}{818} &
  \multirow{-2}{*}{192} &
  \cellcolor[HTML]{EFEFEF}0.73 &
  \multirow{-2}{*}{\textbf{0.81}} &
  \cellcolor[HTML]{EFEFEF}\textbf{0.78} &
  \cellcolor[HTML]{EFEFEF}0 &
  \cellcolor[HTML]{EFEFEF}15 &
  \multirow{-2}{*}{270} &
  \multirow{-2}{*}{740} &
  \cellcolor[HTML]{EFEFEF}0.00 &
  \multirow{-2}{*}{0.27} &
  \cellcolor[HTML]{EFEFEF}0.07 \\ \cline{1-1}
21 &
  107 &
  0 &
  \cellcolor[HTML]{EFEFEF} &
  \cellcolor[HTML]{EFEFEF} &
  \textbf{1.00} &
  \cellcolor[HTML]{EFEFEF} &
  \textbf{0.96} &
  105 &
  2 &
  \cellcolor[HTML]{EFEFEF} &
  \cellcolor[HTML]{EFEFEF} &
  0.98 &
  \cellcolor[HTML]{EFEFEF} &
  0.95 &
  107 &
  0 &
  \cellcolor[HTML]{EFEFEF} &
  \cellcolor[HTML]{EFEFEF} &
  \textbf{1.00} &
  \cellcolor[HTML]{EFEFEF} &
  0.28 \\ \cline{1-1}
22 &
  \cellcolor[HTML]{EFEFEF}68 &
  \cellcolor[HTML]{EFEFEF}0 &
  \multirow{-2}{*}{\cellcolor[HTML]{EFEFEF}340} &
  \multirow{-2}{*}{\cellcolor[HTML]{EFEFEF}269} &
  \cellcolor[HTML]{EFEFEF}\textbf{0.87} &
  \multirow{-2}{*}{\cellcolor[HTML]{EFEFEF}0.56} &
  \cellcolor[HTML]{EFEFEF}\textbf{0.81} &
  \cellcolor[HTML]{EFEFEF}46 &
  \cellcolor[HTML]{EFEFEF}32 &
  \multirow{-2}{*}{\cellcolor[HTML]{EFEFEF}509} &
  \multirow{-2}{*}{\cellcolor[HTML]{EFEFEF}100} &
  \cellcolor[HTML]{EFEFEF}0.59 &
  \multirow{-2}{*}{\cellcolor[HTML]{EFEFEF}\textbf{0.84}} &
  \cellcolor[HTML]{EFEFEF}0.77 &
  \cellcolor[HTML]{EFEFEF}50 &
  \cellcolor[HTML]{EFEFEF}28 &
  \multirow{-2}{*}{\cellcolor[HTML]{EFEFEF}158} &
  \multirow{-2}{*}{\cellcolor[HTML]{EFEFEF}451} &
  \cellcolor[HTML]{EFEFEF}0.64 &
  \multirow{-2}{*}{\cellcolor[HTML]{EFEFEF}0.26} &
  \cellcolor[HTML]{EFEFEF}0.26 \\ \cline{1-1}
23 &
  327 &
  183 &
   &
   &
  \textbf{0.64} &
   &
  \textbf{0.51} &
  94 &
  412 &
   &
   &
  0.19 &
   &
  0.39 &
  17 &
  493 &
   &
   &
  0.03 &
   &
  0.43 \\ \cline{1-1}
24 &
  \cellcolor[HTML]{EFEFEF}74 &
  \cellcolor[HTML]{EFEFEF}89 &
  \multirow{-2}{*}{276} &
  \multirow{-2}{*}{309} &
  \cellcolor[HTML]{EFEFEF}0.45 &
  \multirow{-2}{*}{0.47} &
  \cellcolor[HTML]{EFEFEF}\textbf{0.54} &
  \cellcolor[HTML]{EFEFEF}54 &
  \cellcolor[HTML]{EFEFEF}109 &
  \multirow{-2}{*}{408} &
  \multirow{-2}{*}{177} &
  \cellcolor[HTML]{EFEFEF}0.33 &
  \multirow{-2}{*}{\textbf{0.70}} &
  \cellcolor[HTML]{EFEFEF}0.53 &
  \cellcolor[HTML]{EFEFEF}163 &
  \cellcolor[HTML]{EFEFEF}0 &
  \multirow{-2}{*}{240} &
  \multirow{-2}{*}{345} &
  \cellcolor[HTML]{EFEFEF}\textbf{1.00} &
  \multirow{-2}{*}{0.41} &
  \cellcolor[HTML]{EFEFEF}0.47 \\ \cline{1-1}
25 &
  13 &
  0 &
  \cellcolor[HTML]{EFEFEF}124 &
  \cellcolor[HTML]{EFEFEF}138 &
  \textbf{1.00} &
  0.47 &
  \textbf{0.93} &
  13 &
  0 &
  \cellcolor[HTML]{EFEFEF}194 &
  \cellcolor[HTML]{EFEFEF}68 &
  \textbf{1.00} &
  \textbf{0.74} &
  \textbf{0.93} &
  0 &
  13 &
  \cellcolor[HTML]{EFEFEF}102 &
  \cellcolor[HTML]{EFEFEF}160 &
  0.00 &
  0.39 &
  0.21 \\ \hline
Avg. &
  - &
  - &
  - &
  - &
  \textbf{0.73} &
  \textbf{0.69} &
  \textbf{0.75} &
  - &
  - &
  - &
  - &
  0.65 &
  0.52 &
  0.56 &
  - &
  - &
  - &
  - &
  0.44 &
  0.52 &
  0.35 \\ \hline
\end{tabular}%
}
}
\caption{Evaluation Result. TP, FP, TN and FN stand for True/False Positive, True/False Negative. We use common definition of metrics. $TPR = TP/(TP+FN)$ and $TNR = TN/(TN+FP)$ and $AUR.$ is Area Under ROC curve.} 
\label{table:result-main}
\end{table*}

\noindent\textbf{Pretraining Set.} 
The {\tt DARPA TRACE} dataset provides an extensive volume of logs collected from various tracepoints. We begin by excluding the portions of the trace used to construct the attack and benign sets. Pretraining is then conducted separately for each platform (i.e., Windows and Linux), resulting in two distinct models. The pretraining set includes five different tracepoints for Linux, spanning 12, 10, 10, 10, and 10 days (4.3 TB of uncompressed logs), respectively, and four tracepoints for Windows, spanning 8, 10, 10, and 10 days (3TB), respectively. To the best of our knowledge, \dupin is the state-of-the-art pretrained models, trained on the most extensive volume of provenance logs. We also use this dataset for building the tokenizer. Details on {\tt DARPA TRACE} dataset can be found in Appendix~\ref{subsec:usage-darpa-dataset}. We note that later-produced audit logs may have used for pre-training the base model and early-produced logs have been used for evaluation in some cases. In all cases, our pretraining does not temporally overlap with attack logs.

\subsection{Experiment Setup}\label{sec:experiment-setup}

\noindent\textbf{Implementation.} The core implementation of \dupin is wrapped up as our toolkit, consisting of the audit parser (2,200 lines), graph builder (500 lines). See our artifact repo for more details about our toolkit. It is compatible with three popular audit formats: \textit{Linux Audit Framework} (500 lines), \textit{Common Data Model} (\textit{CDM}, 1000 lines)~\cite{khoury2020event}, and {\tt Windows ETW~\cite{windowsetw}} (700 lines). On top of the toolkit, \dupin implements a GNN module (500 lines), tokenizer (500 lines). 

For GNNs, we use \textit{PyTorch}~\cite{pytorch} and \textit{DGL}~\cite{dgl} library. We have evaluated \dupin{} on a \texttt{Ubuntu} 22.04, equipped with a 3.00GHz \texttt{Xeon} CPU, 256GB memory and \texttt{RTX} \texttt{A6000} GPU. 

\noindent\textbf{Parameters.} The learning rate is set to $10^{-4}$ for pretraining (mask prediction). Afterwards, it is adjusted to $10^{-5}$ to preserve the pre-trained acquirement when these cells are fine-tuned. Meanwhile, the learning rate for additional classifier layers is higher as $10^{-4}$ to promote the attack learning.

Each embedding size is set to 128 which is similar size to word embedding. The base model is 7 GAT layers. The classifier is fully-connected with one hidden layer of 20 cells. For the masked learning, we randomly select 25\% of nodes and edges. We run one epoch for pretraining and 200 epochs for fine-tuning. For the tokenizer, we set $\tau = 20K$ (for {\tt Linux}) and $30K$ (for {\tt Windows}). As aforementioned, we set $\gamma = 40$, $\delta = 100$ for subgraph discovery.

\noindent\textbf{Evaluation Metrics.} We use \textit{true positive rate} (TPR), \textit{true negative rate} (TNR), and \textit{area under ROC} (AUROC) for our metrics. Additionally, the numbers of \textit{true positives} (TP), \textit{true negatives} (TN), \textit{false positives} (FP), and \textit{false negatives} (FN) are provided in Appendix. Note that \textit{precision} and \textit{recall} can be misleading when the validation set is biased. For example, a very small error in false positives can lead to a very low recall rate due to the dominating size of the benign set.

\noindent\textbf{Threshold.} We determine the threshold that maximizes the equally valuing both TPR and TNR. Note that if we use the accuracy as threshold search heuristic, true positives can be overlooked due to data imbalance. To this end, \dupin sets the threshold where it minimizes the Euclidean distance to the perfect spot (i.e., the point in the ROC curve closest to the one with TPR=$1.0$, TNR=$1.0$). The threshold is determined separately for each model during training. Note that implementation details can be found in our code repository.

\subsection{Research Questions} 

To systematically evaluate \dupin, we answer the following research questions. 

\noindent\textbf{RQ1: How does \dupin compare with other self-supervised, pre-trainable models?} We use two self-supervised learning based methods, \texttt{AirTag}~\cite{ding2023airtag} and \texttt{MAGIC}~\cite{jia2024magic} to answer this research question. \texttt{AirTag} employs text-level analysis with the BERT model to obtain contextualized embeddings. After transforming logs into embeddings, it applies a one-class learning algorithm to detect outliers. \texttt{MAGIC} uses graph-based approach. 
AirTag (masked language modeling) and MAGIC (contrastive learning) employ pre-training as a workflow, and were therefore selected as our primary baselines. Those two baselines (\texttt{AirTag}~\cite{ding2023airtag} and \texttt{MAGIC}~\cite{jia2024magic} enables the direct comparison between unsupervised learning and our few-shot method with minimal architecture differences. We further compare \dupin with various approaches in Section~\ref{sec:ablation-study}.

\noindent\textbf{RQ2: How does \dupin compare with other approaches with different GNN architectures?} Here, we evaluate \dupin against alternative model architectures. First, we include the unsupervised learning approach, \texttt{Flash}~\cite{rehman2024flash}, which incorporates temporal ordering features and word embeddings on its graph networks. We also include \texttt{Kairos}~\cite{cheng2024kairos}, which models temporal context by constructing graph queues over sliding time windows.
To evaluate the performance gain by our attentive model (i.e., \texttt{GAT}), we evaluate \dupin against different GNN models - the graph isomorphism network ({\tt GIN})~\cite{xu2018powerful} and the graph convolutional network ({\tt GCN})~\cite{kipf2016semi} and the most recent baseline~\cite{bilot2025sometimes}.

\noindent\textbf{RQ3: (Ablation Study) How much do pre-training and few-shot learning contribute to the performance?} 
First, we ablate the pretraining phase in \dupin’s learning pipeline, shifting it to rely entirely on few-shot learning, in order to quantify the performance gains contributed by large-scale pretraining.
We also evaluate the performance of \dupin under a minimal number of few-shot training samples, by limiting the size of attack learning. This evaluation highlights the minimal number of few-shot samples necessary for \dupin{}.

\noindent\textbf{RQ4: What is the time complexity of \dupin?}
We measure the time per log for each component of \dupin (i.e., log processing, graph construction and model training/inference).

\subsection{RQ1: Effectiveness of \dupin against other self-supervised pre-trained models}\label{sec:eval-effectiveness}

\noindent\textbf{Training.}
To adhere to our evaluation principle, each model is used for attack sets that were not used during training. Specifically, we train a separate model for each dataset source: (i) \texttt{DARPA TRACE} ({\tt Linux}), (ii) \texttt{DARPA TRACE} ({\tt Windows})~\cite{darpa2014trust}, (iii) \texttt{ATLAS} ({\tt Windows})~\cite{alsaheel2021atlas}, and (iv) \texttt{PalanTir} ({\tt Linux})~\cite{zeng2022palantir}, then perform cross-validation using different sources. Note that we build separate tokenizers and pre-trained models for each O/S (i.e., \texttt{Linux} or \texttt{Windows}). Cross-validation is therefore performed separately within the same platform: (1) between \texttt{DARPA TRACE - Linux} dataset (ID: 1--7) and \texttt{PalanTir} dataset (ID: 20--25), and (2) between \texttt{DARPA TRACE - Windows}  (ID: 8--12) and \texttt{ATLAS} dataset (ID: 13--18). For example, an attack set from \texttt{DARPA TRACE - Linux} will be evaluated with model trained under \texttt{PalanTir} dataset. As a result, each model is trained on only five to seven attack sets, reflecting the limited availability of attack training data. As \texttt{AirTag} uses textual logs, we slightly modify version of Algorithm~\ref{alg:subgraph-discovery} to follow their methodology, thereby we use the log trace of chosen node (at line 6), instead of leveraging subgraph discovery. We have verified that all attack sets include the corresponding attack events. \texttt{MAGIC} utilizes graph representation learning, followed by outlier detection as a downstream task. Both methods use self-supervised learning and outperform existing learning-based methodologies.

\noindent\textbf{Effectiveness.} Table~\ref{table:result-main} presents the evaluation results. In this table, attack IDs correspond to those in Table~\ref{table:dataset-summary} (i.e., attack campaign IDs). Note that \dupin operates at graph-level granularity (Section~\ref{sec:design-graph-construction} and Appendix~\ref{subsec:subgraph-discovery}). The number of entities per attack campaign, therefore, represents the subgraphs extracted from audit logs.

We observe that \dupin{} outperforms all baseline methods in terms of TPR, TNR and AUROC. Overall, it achieves a 21\% improvement in AUROC over the best-performing baseline (as shown in the last row). Specifically, \dupin{} yields an 8\% higher TPR and a 17\% higher TNR. Furthermore, our method maintains the best performance in 19 out of 25 attack sets, demonstrating supremacy in consistency. Moreover, all four dataset sources consistently demonstrate the superiority of \dupin{}. \dupin{} achieves the highest true positives in 15 attack sets. This is critical for reconstructing a complete attack story, as it reduces the risk of missing entities or disruptions in the sequence of attack events. Additionally, the 17\% gain in TNR brings us one step closer to automation by reducing the human effort required to investigate false alarms.

Note that relative performance tend to decrease when applying our novel cross-validation methodologies. We have evaluated \dupin{} under a relaxed cross-validation setup—where training and testing are conducted within the same dataset source: (i) \texttt{DARPA TRACE - Linux} (ID: 1--7), (ii) \texttt{PalanTir} (ID: 20--25), (iii) \texttt{DARPA TRACE - Windows} (ID: 8--12), (iv) \texttt{ATLAS} (ID: 13--18). For each attack set, a model trained on the remaining attack sets from the same dataset source is used (e.g., For attack ID 1, the model trained attacks IDs from 2 to 7).  Under such settings, \dupin achieves a TPR of 86\% and a TNR of 80\% (see Appendix \ref{subsec:relaxed-corss-val} for details). 

\begin{figure*}[]
    \centering
    {
    \includegraphics[clip=true,trim=0mm 0mm 0mm 0mm,width=1.0\linewidth]{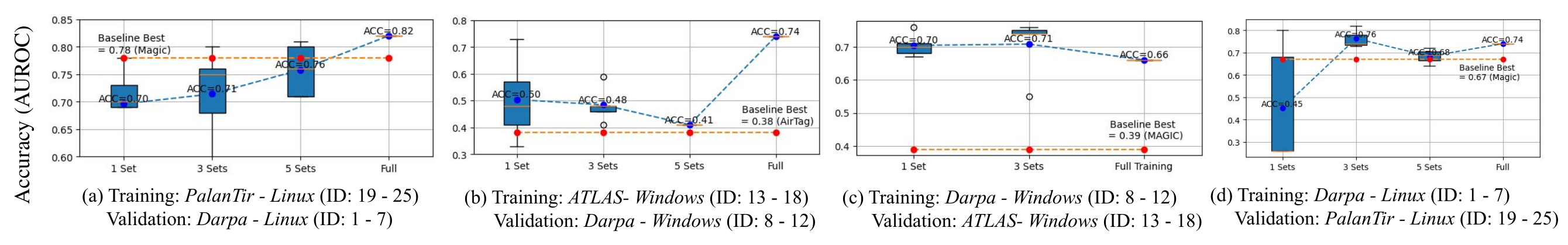}
    }
    \vspace{-3mm}
    \caption{\centering Evaluation on accuracy impact of training size limit. Each X-axis refers to the reduced training size, i.e., the number of attack sets to train each model. At each training size, we conduct 5 times of random sampling for partial training subsets and draw box-plot (blue squares) on that axis to denote their accuracies. Original trainings of \dupin are also included (marked ``Full'') in each graph. Note that original training sizes can be found in Table~\ref{table:dataset-summary}. The best accuracy among baselines (i.e., \texttt{AirTag} or \texttt{MAGIC}) is also marked (red dashed line) for comparison.}
	\label{fig:ablation-study-fine-tune}
\end{figure*}

\subsection{RQ2: Effectiveness of \dupin against other GNN architectures}\label{sec:eval-effectiveness-2}
\noindent\textbf{\dupin against Model with Temporal Features.}
\texttt{Flash}~\cite{rehman2024flash} proposes a novel graph architecture, using temporal features and word-level embedding. To enable a fair comparison with \texttt{Flash}, we integrate its implementation into our training and evaluation pipeline using the same datasets. Accordingly, we modify the data loader component in the \texttt{Flash} implementation to accommodate our datasets. We provide the event order from audit logs accordingly. 
\texttt{Kairos}~\cite{cheng2024kairos} applies temporal encoding over sliding time windows and detects anomalies by scoring reconstruction errors. We use the original source code of \texttt{Kairos} to construct graph queues over time windows, feeding audit logs with event timestamp information from our datasets.

\begin{table}[H]
\resizebox{\columnwidth}{!}{
\begin{tabular}{|ccc|ccc|ccc|}
\hline
\multicolumn{3}{|c|}{\texttt{Flash}} & \multicolumn{3}{c|}{\texttt{Kairos}} & \multicolumn{3}{c|}{\dupin}                   \\ \hline
\textbf{TPR} & \textbf{TNR} & \textbf{AUR.} & \textbf{TPR}  & \textbf{TNR} & \textbf{AUR} & \textbf{TPR} & \textbf{TNR}  & \textbf{AUR}  \\ \hline
0.60         & 0.51         & 0.55          & 0.57 & 0.52         & 0.51         & \textbf{0.73}         & \textbf{0.69} & \textbf{0.75} \\ \hline
\end{tabular}
}
\caption{\centering Results on Models using temporal feature (i.e., \texttt{Flash}, \texttt{Kairos}) and \dupin}
\label{table:comparison-flash-kairos}
\vspace{-2mm}
\end{table}

Table~\ref{table:comparison-flash-kairos} presents the results of our comparative evaluation. While incorporating temporal features enhances performance (for example, \texttt{Flash} achieves AUROC scores comparable to \texttt{MAGIC} without requiring pretraining), \dupin{} still outperforms  baseline methods, achieving the highest overall accuracy. The subdivided accuracies over all datasets are included in Appendix~\ref{subsec:full-flash-kairos-gin-gcn}

\noindent\textbf{\dupin against Different Graph Neural Networks.}
We compare \dupin against different GNN models: the graph isomorphism network ({\tt GIN}) and the graph convolutional network ({\tt GCN}).
For this comparison, we replace the compatible GNNs with our base GAT component, while keeping all other components unchanged.

\begin{table}[H]
\resizebox{\columnwidth}{!}{
\begin{tabular}{|ccc|ccc|ccc|}
\hline
\multicolumn{3}{|c|}{\textbf{GIN}}                   & \multicolumn{3}{c|}{\textbf{GCN}}                    & \multicolumn{3}{c|}{\dupin}                   \\ \hline
\textbf{TPR} & \textbf{TNR} & \textbf{AUR.} & \textbf{TPR}  & \textbf{TNR} & \textbf{AUR} & \textbf{TPR} & \textbf{TNR}  & \textbf{AUR}  \\ \hline
0.51         & 0.64         & 0.57          & \textbf{0.77} & 0.56         & 0.70         & 0.73         & \textbf{0.69} & \textbf{0.75} \\ \hline
\end{tabular}
}
\caption{Results on Non-Attentive Models and \dupin}
\label{table:abalation-no-attentive}
\vspace{-2mm}
\end{table}

Table~\ref{table:abalation-no-attentive} presents the results. We observe that \dupin outperforms the two non-attentive GNNs, achieving a 5\% AUROC improvement over {\tt GCN}. The {\tt GIN} model exhibits higher false negatives (TPR = 0.51), whereas {\tt GCN} struggles with higher false positives (TNR = 0.56). In contrast, \dupin demonstrates comparatively balanced TPR and TNR, leading to better overall performance in AUROC. The findings demonstrate the necessity of attention component. The entire results for whole attack campaigns are provided in Appendix~\ref{subsec:full-flash-kairos-gin-gcn}.

\noindent\textbf{Comparing \dupin with Additional Baselines.} 
We further compare \dupin with \texttt{VELOX}~\cite{bilot2025sometimes} and \texttt{TAPAS}~\cite{zhang2025tapas}, the most recent baselines available at the time of publication. The key insight of \texttt{VELOX} is that complex architectures can paradoxically degrade detection performance. It instead proposes simple network structures paired with \texttt{Word2Vec} embeddings. \texttt{TAPAS} is another recently published work which reduces the analysis complexity by eliminating sparsity in provenance data. 

\begin{table}[H]
\centering
\resizebox{\columnwidth}{!}{
\begin{tabular}{|ccc|ccc|ccc|}
\hline
\multicolumn{3}{|c|}{\texttt{VELOX}} & \multicolumn{3}{c|}{\texttt{TAPAS}} & \multicolumn{3}{c|}{\textbf{\dupin}}                   \\ \hline
\textbf{TPR} & \textbf{TNR} & \textbf{AUR.} & \textbf{TPR}  & \textbf{TNR} & \textbf{AUR} & \textbf{TPR} & \textbf{TNR}  & \textbf{AUR}  \\ \hline
0.54 & 
0.60 & 
0.58 & 
0.54 & 
0.51 & 
0.51 & 
\textbf{0.73} & 
\textbf{0.69} & 
\textbf{0.75} \\ \hline
\end{tabular}
}
\caption{Results on \texttt{VELOX} and \texttt{TAPAS}}
\label{table:dupin-vs-velox}
\vspace{-2mm}
\end{table}

Table~\ref{table:dupin-vs-velox} shows the results. As a relatively recent work, \texttt{VELOX} outperforms all existing baselines using unsupervised methods (i.e., \texttt{Magic}~\cite{jia2024magic}, \texttt{AirTag}~\cite{ding2023airtag} \texttt{Flash}~\cite{rehman2024flash} and \texttt{Kairos}~\cite{cheng2024kairos}). The accuracy of \texttt{TAPAS} remains similar to earlier baselines. However, the comparative results still shows the need of \dupin's attack learning, outperforming \texttt{VELOX} by 17\% of AUROC. The entire results for whole attack campaigns are provided in Appendix~\ref{subsec:full-flash-kairos-gin-gcn}.

\subsection{RQ3: Ablation Study}\label{sec:ablation-study}

\noindent\textbf{\dupin without Pre-training.}
We examine the impact of pretraining on model performance. Hence, as a baseline, we use an ablated version of \dupin that omits the pretraining phase and instead relies on randomly initialized scratch models. The evaluation follows the same configuration used in Table~\ref{table:result-main}. Table~\ref{table:abalation-no-pretraining} shows the results. Overall, the pretrained \dupin outperforms the baseline with a 12\% improvement in TPR, while also achieving a 4\% gain in TNR.
The detailed results on this evaluation can be found in Appendix~\ref{subsec:result-ablation}.

\begin{table}[H]
\centering
\resizebox{\columnwidth}{!}{
\begin{tabular}{|ccc|ccc|ccc|}
\hline
\multicolumn{3}{|c|}{\textbf{\dupin w/ pretrain}} & \multicolumn{3}{c|}{\textbf{\dupin wo/ pretrain}} & \multicolumn{3}{c|}{\textbf{Difference}}                   \\ \hline
\textbf{TPR} & \textbf{TNR} & \textbf{AUR.} & \textbf{TPR}  & \textbf{TNR} & \textbf{AUR} & \textbf{TPR} & \textbf{TNR}  & \textbf{AUR}  \\ \hline
\textbf{0.73} & 
\textbf{0.69} & 
\textbf{0.75} & 
0.61 & 
0.65 & 
0.59 & 
\textbf{+ 0.12} & 
\textbf{+ 0.04} & 
\textbf{+ 0.16} \\ \hline
\end{tabular}
}
\caption{\dupin w/ Pretraining vs wo/ Non-Pretraining}
\label{table:abalation-no-pretraining}
\vspace{-2mm}
\end{table}

Despite its best extensive coverage on experimental datasets, we believe that \dupin still holds significant potential for improvement through greater diversity and scale in training by an expectation supported by the recent success of large-scale models~\cite{achiam2023gpt}. In real-world or enterprise-level settings, we anticipate the feasibility of large-scale pretraining on more diverse tracepoints and extended trace durations.

\noindent\textbf{\dupin on Limited Attack Fine-Tuning.}
In this study, we evaluate \dupin under limited fine-tuning training. Originally, each model of \dupin is trained with 5 to 7 attack sets (Table~\ref{table:dataset-summary}). We further reduce the training size to compare \dupin with the baselines when trained under fewer attack sets. We train \dupin with different reduced combinations of attack sets, while keeping the validation sets unchanged. Note that our original evaluation involves training four models: (i) {\tt Darpa - Linux}, (ii) {\tt Darpa - Windows}, (iii) {\tt ATLAS}, and (iv) {\tt PalanTir}. For each model, we randomly extract combinations of 1, 3, and 5 sets. The random extraction is repeated five times.

Figure~\ref{fig:ablation-study-fine-tune} presents the AUROC scores of \dupin under reduced training. In three cases ((b), (c) and (d)), \dupin achieves superior performance over the best baseline when trained with three attack sets. Moreover, in cases (b) and (c), \dupin trained with only a single attack set still outperforms baselines. Our findings indicate that \dupin (i.e., pretraining, few-shot learning) can serve as a better alternative to baselines (i.e., one-class outlier detection), even when the available attack training data is limited to as few as 3 to 5 APT sets.

We further measure the variance arising from different few-shot sample selections. In our original evaluations, each model is trained on 5–7 attack campaigns. For each model, we then sample three sets for train and measure the variance across all of combinations. 

\begin{table}[H]
\small
\begin{tabular}{|c|cccc|}
\hline
\multicolumn{1}{|c|}{\textbf{Training}} & \multicolumn{1}{c|}{\textbf{Mean}} & \multicolumn{1}{c|}{\textbf{Min/Max}} & \multicolumn{1}{c|}{\textbf{S.D.}} & \textbf{Baseline}              \\ \hline
\texttt{Darpa - Windows}                      & \cellcolor[HTML]{EFEFEF}0.70     & \cellcolor[HTML]{EFEFEF}0.67/0.76        & \cellcolor[HTML]{EFEFEF}0.04     & \cellcolor[HTML]{EFEFEF}0.39 \\ \cline{1-1}
\texttt{Darpa - Linux}                      & 0.76                             & 0.73/0.82                                 & 0.05                             & 0.67                        \\ \cline{1-1}
\texttt{Atlas - Windows}                        & \cellcolor[HTML]{EFEFEF}0.48     & \cellcolor[HTML]{EFEFEF}0.41/0.58         & \cellcolor[HTML]{EFEFEF}0.07     & \cellcolor[HTML]{EFEFEF}0.38 \\ \cline{1-1}
\texttt{PalanTir - Linux}                             & 0.71                             & 0.58/0.80                               & 0.09                             & 0.78                        \\ \hline
\end{tabular}
\caption{Variances of \dupin's score (i.e., AUROC) with Reduced Few Shot Train. Each model is trained on combination of three attack campaigns. We evaluate the variances over different combinations. S.D. stands for standard deviation.}
\label{table:variance-test}
\vspace{-3mm}
\end{table}

Table~\ref{table:variance-test} shows the results.
Among the models that outperform the baselines (except the model trained with \texttt{PalanTir-Linux}) consistently remain above the baselines regardless of the specific choice of few-shot samples. \dupin may exhibit accuracy deviations within $(max-min)/2$ (0.045 -- 0.11) or standard deviations (0.04 -- 0.09). It can be further reduced with larger few-shot samples (each \dupin model learns from 5 to 7 sets).

\subsection{RQ4: Efficiency of \dupin}
\dupin{} incurs both (i) the runtime which is related to the real-time surveillance such as log processing and attack detection and (ii) offline tasks needed for model training. In this section, we measure efficiency in both tasks. We include two different tracepoints from the \texttt{Darpa TRACE}, \texttt{Cadets, Trace} and trim them to represent a day trace (24 hours each). The specification of our resources is stated in Section~\ref{sec:experiment-setup}. We utilize a single GPU and 15 of batch size. The processing time is ratio of elapsed time from the given task over 24 hours.

\noindent\textbf{Runtime Cost.} To facilitate runtime surveillance, \dupin{} must (i) load and parse the text-based audit logs, converting them into data structures, (ii) construct provenance graphs, which involves extensive search complexity, and (iii) execute the GAT model. Accordingly, we assess the time by breaking down each task. 

Table~\ref{table:runtime-cost} shows the results. The accumulated runtime rate is 25.58\%, 15.55\%, respectively. In both cases, the logs are processed faster than they accumulate. We observe that the most bottlenecking task of \dupin{} is the data processing - specifically, log parsing and graph construction - rather than the significant components of proposed methods, such as GAT-based models. This is a common observation in many cyber forensics techniques. As a remedy, one may apply optimized sequential models such as \textit{Mamba}~\cite{gu2023mamba}. However, optimization of sequence processing is beyond the scope of this paper. 

\begin{table}[H]
\small
\begin{tabular}{|c|cccc|}
\hline
                                 & \multicolumn{2}{c|}{\textit{Trace}}                                         & \multicolumn{2}{c|}{\textit{Cadets}}                                \\ \cline{2-5} 
\multirow{-2}{*}{\textbf{Tasks}} & \multicolumn{1}{c|}{\textbf{Time}} & \multicolumn{1}{c|}{\textbf{Rate}} & \multicolumn{1}{c|}{\textbf{Time}} & \textbf{Rate}              \\ \hline
Log Parsing                      & \cellcolor[HTML]{EFEFEF}04h19m     & \cellcolor[HTML]{EFEFEF}18.03\%        & \cellcolor[HTML]{EFEFEF}01h09m     & \cellcolor[HTML]{EFEFEF}4.81\% \\ \cline{1-1}
Graph Const.                     & 01h40m                             & 6.98\%                                 & 02h34m                             & 10.73\%                        \\ \cline{1-1}
GAT model                       & \cellcolor[HTML]{EFEFEF}00h08m     & \cellcolor[HTML]{EFEFEF}0.57\%         & \cellcolor[HTML]{EFEFEF}00h01m     & \cellcolor[HTML]{EFEFEF}0.01\% \\ \cline{1-1}
Total                            & 06h08m                             & 25.58\%                                & 03h44m                             & 15.55\%                        \\ \hline
\end{tabular}
\caption{Runtime  Cost of \dupin{} on 24-hour Logs.}
\label{table:runtime-cost}
\vspace{-3mm}
\end{table}

\noindent\textbf{Time Cost on Training.} The training of \dupin{} involves masked learning (i.e., pretraining) and fine-tuning (i.e., attack classifier). We measure the elapsed time for training in both tasks. The results are denoted in Table~\ref{table:runtime-training}. 

\begin{table}[H]
\small
\begin{tabular}{|c|cccc|}
\hline
                                 & \multicolumn{2}{c|}{\textit{Trace}}                                         & \multicolumn{2}{c|}{\textit{Cadets}}                                \\ \cline{2-5} 
\multirow{-2}{*}{\textbf{Tasks}} & \multicolumn{1}{c|}{\textbf{Time}} & \multicolumn{1}{c|}{\textbf{Rate}} & \multicolumn{1}{c|}{\textbf{Time}} & \textbf{Rate}              \\ \hline
Pre-training                  & \cellcolor[HTML]{EFEFEF}06m27s     & \cellcolor[HTML]{EFEFEF}0.24\%         & \cellcolor[HTML]{EFEFEF}03m27s     & \cellcolor[HTML]{EFEFEF}0.38\% \\ \cline{1-1}
Fine-Tuning                      & 19m22s                             & 1.35\%                                 & 85m09s                          & 5.94\%                         \\ \hline
\end{tabular}
\caption{Runtime  Cost of \dupin{} on Training.}
\label{table:runtime-training}
\vspace{-3mm}
\end{table}

The cost of masked learning is an runtime rate of 0.24\% and 0.38\% which is comparable to that of the inference task (0.57\%, 0.01\%). It implies that masked learning does not introduce a significant amount of additional complexity over our base model. While the training time can increase linearly with the size of the pretraining data (as pursuing a large pretrained model), pretraining is a one-time expense in practice, alleviating such concerns about computational costs. Moreover, well-pretrained models can be shared and reused.

\subsection{Case Study}
\label{sec:case-study}

We present two representative cases for case study. Both attacks use common attack techniques, including compromise and backdoor installation. 

\noindent\textbf{Attack Case 1.} 
Set M-1 (Attack ID: 13): The victim user accessed a malicious host ({\tt index.html}) via the {\tt Firefox} browser. Then, the backdoor file ({\tt pypayload.exe}) is installed which further executes their operation by executing the {\tt cmd.exe}.

\dupin{} successfully detects this web-based compromise, whereas both baseline models fail to raise any alerts. Notably, the context of backdoor installation is exhibited in various forms. For example, a {\tt .zip} file is downloaded and extracted into an executable (ID: 12). In another case, we observe {\tt powershell} accessing external IP addresses to download exploit-laden {\tt word} documents (ID: 10). Similarly, {\tt firefox} is seen writing a {\tt .rtf} file used to compromise the system (ID: 8).

\noindent\textbf{Attack Case 2.} 
Linux Backdoor (Attack ID: 1): This attack comprises multiple steps of malicious activity. First, the file {\tt vUgefal} is downloaded by an {\tt nginx} server. Subsequently, the attacker gathers system information, such as by accessing {\tt /etc/passwd}.

The \dupin model for {\tt Linux}, trained on attack sets 19 through 25, successfully detects this attack. Notably, the training data includes a variant involving the download of a {\tt .sh} backdoor using {\tt wget} (ID: 19). It also covers various data leakage attempts, such as the exfiltration of {\tt .txt} files (IDs: 20, 23) and access to sensitive files like {\tt /etc/group} and {\tt /etc/passwd} (ID: 22). \dupin effectively detects the current attack, leveraging its exposure to these related patterns.

\noindent\textbf{Implications.} 
While both cases demonstrate the use of similar attack techniques, they exhibit different manifestations such as distinct symbols or victim programs between the training phase and inference. This demonstrates \dupin's ability to generalize learned attack patterns beyond exact matches.

\noindent\textbf{Failing Case 1.} While \dupin outperforms the baselines in most cases, it falls behind \texttt{AirTag} on attack sets ID 9 and 12. We illustrate this issue through a case study of one such instance (ID: 12). The attack begins with a compromised instance of {\tt firefox}, initiated by access from an external IP address ({\tt 135.84.161.202}). Subsequently, a backdoor file named {\tt hJauWl01} is downloaded to the local directory, where the attacker attempts to elevate privileges. 

\begin{figure}[]
    \centering
    {%
    \includegraphics[clip=true,trim=0mm 0mm 0mm 0mm,width=0.9\linewidth]{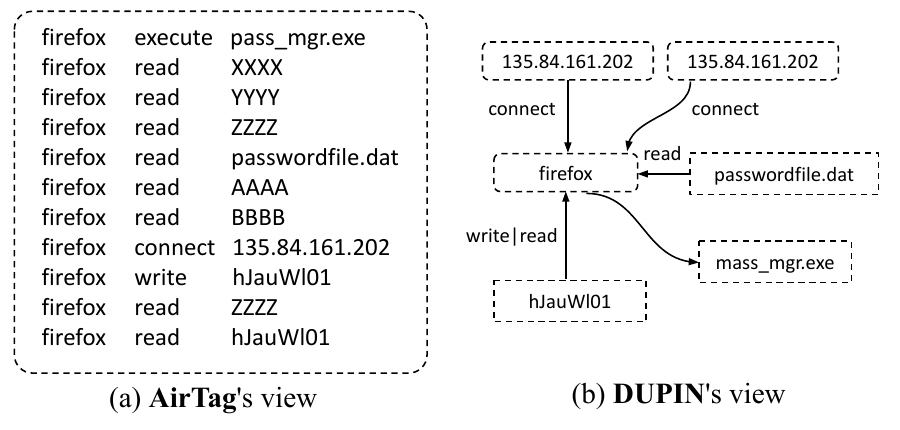}
    }%
	\caption{Text Logs vs Graph Representation on Attack 12}
	\label{fig:case-study-failing}
\end{figure}

The performance of \dupin is degraded in this case due to graph-based structures. While this representation may limit effectiveness in certain scenarios, graphs are generally well-suited for modeling attack provenance. This is evidenced by the overall performance gains observed with both \dupin and \texttt{MAGIC}, which leverage graph-based approaches. However, in this specific attack, the graph representation fails to preserve the time order of events - an aspect that remains intact in the original textual logs. Figure~\ref{fig:case-study-failing} shows how each method captures the attack provenance. 

In the textual logs, the download of {\tt hJauWl01} is immediately followed by a connection from the external server ({\tt 135.84.161.202}), which could raise the suspicion level of the model. However, this time-based information is lost in the graph representation, reducing the model’s confidence in detecting this attack. While this issue could be addressed by the temporal ordering embeddings~\cite{rehman2024flash,cheng2024kairos}, we leave it out of scope. In this paper, we primarily focus on few-shot attack learning with an extensively pretrained model.

\noindent\textbf{Failing Case 2.} We present cases in which the observable attack evidences are relatively weak, arising from execution failures or ambiguity in the attacker’s objectives.

Regarding the phishing campaign (ID: 5), \dupin, along with all baselines, fails to detect any of the attacks. In this campaign, the attacker tries to exploit a vulnerable e-mail client ({\tt pine}); however, according to the official {\tt Darpa} documentation, {\tt pine} was no longer running due to a system restart at the intrusion time. Our manual inspection also confirms that such logs do not exist. As a result, the attack attempts yield only fragmented traces, consisting mainly of several failed connection attempts to {\tt bash} and the appearance of a pre-installed backdoor file ({\tt tcexec}). Because these behaviors are carried out separately across different attack trails rather than forming a coherent attack plan, the traces of this attack become fragmented into different model input segments.

In other case (ID: 6), the attack objective is weakly manifested. Our manual inspection confirms that the activities involve an execution of {\tt perl} scripts and an establishment of C2 connections, but without further progress. \dupin does not raise its suspicious score enough to meet the threshold.

\textbf{Use Case.} We demonstrate a use case of \dupin, illustrating the analyst's use of its analysis results. We use attack case from \texttt{Darpa Trace}, \textit{Windows} host against \texttt{Firefox} (attack ID 8). In this attack campaign, the attacker compromised \texttt{Firefox} to attain an access and exfiltrated files on the compromised system.
\dupin performs a postmortem attack analysis. It starts upon a notification of attack (e.g., the detection of IoC by firewall).
Then, the analyst runs \dupin with the raw audit logs from the compromised host.

\begin{figure}[]
    \centering
    {%
    \includegraphics[clip=true,trim=0mm 0mm 0mm 0mm,width=0.85\linewidth]{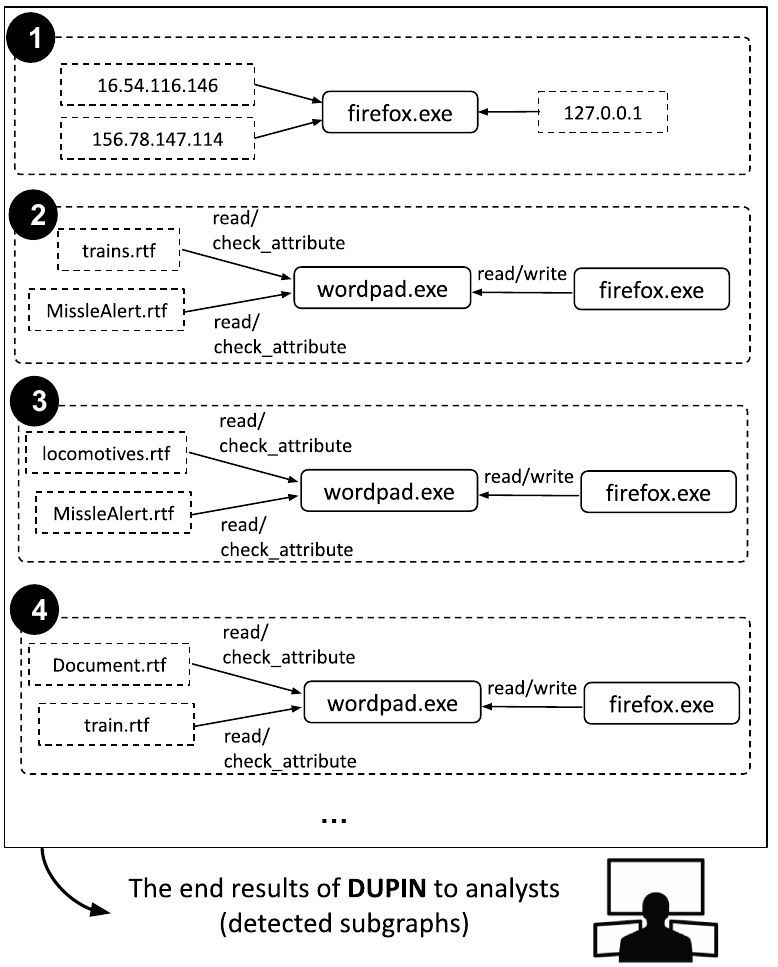}
    }%
	\caption{The end results of \dupin from the raw audit logs}
	\label{fig:use-case-example}
\end{figure}

Figure~\ref{fig:use-case-example} shows the end results of \dupin. 
Since the detection granularity is graph, \dupin generates subgraphs that highlight the attack chains. First, the analyst confirms the initial compromise vectors (i.e., \texttt{16.54.116.146}, \texttt{158.78.147.114} and \texttt{Firefox}) from \circnum{1}. The analyst can immediately block the access from those network origins and guide user to update the program to a recent version.
Also, \dupin highlights the continuing attack behaviors, allowing to identify the leakage of document files (\circnum{2}--\circnum{4}). Then, analyst may reference our results to write an attack report.

Overall, the purpose of \dupin lies in automation of forensic analysis, significantly reducing the attack response time. 

\section{Related Work}
\noindent\textbf{Attack Forensics.} Since the advent of auditing systems, mainstream research has focused on scalability issues. In this regard, many researchers have proposed the compression algorithms~\cite{lee2013loggc,tang2018nodemerge,xu2016high,xie2012hybrid,xie2011compressing} and audit data reductions~\cite{lee2013high,ma2017mpi,ji2018enabling}. Owing to these efforts and advanced computational resources, audit analysis has attained the improved feasibility in reality. Consequently, following research interests have focused on the effectiveness of attack forensics rather than its efficiency.

First, researchers have sought additional information to construct semantically improved graphs.
For example, some work~\cite{ma2016protracer,ji2018enabling} leverages the instrumental information, utilizing code analysis. Additionally, various sources of user-level information such as UI tracking~\cite{yang2020uiscope} and application logs~\cite{yu2021alchemist} have been also utilized. 

Meanwhile, the mainstream research has been challenged to develop algorithmic or learning-based systems that can reduce or replace human efforts. In response, early-stage research to reduce forensics effort has focused on bridging the semantic gaps between low-level system artifacts and human-level threat knowledge. For example, \textit{HOLMES}~\cite{milajerdi2019holmes} defines rule-based prerequisites for representative attack techniques, enabling the high-level behavior summarization. Also, \textit{WATSON}~\cite{zeng2021watson} uses clustering algorithms to construct behavior abstracted embeddings. 

Recently, the trend has shifted towards data-driven approaches~\cite{du2017deeplog,han2020unicorn,alsaheel2021atlas,han2021sigl,zengy2022shadewatcher,dong2023distdet,yang2023prographer,cheng2024kairos} with the growing potential of machine learning applications. This work includes various neural architectures such as sequential models~\cite{du2017deeplog,alsaheel2021atlas}, graph learning~\cite{zengy2022shadewatcher,yang2023prographer,cheng2024kairos} and behavioral embedding~\cite{yang2023prographer}. However, this approach faces a major limitation due to dataset imbalance. To address this problem, lately proposed methods resort to self-supervised learning~\cite{ding2023airtag,jia2024magic} (our baselines), achieving state-of-the-art results. 
In contrast to the above lines of work, \dupin is the first system to combine large-scale self-supervised pretraining on benign provenance with few-shot supervised adaptation on a small attack corpus.

\noindent\textbf{Graph Learning.} Some specific domains require graph-structured data, such as social networks~\cite{sharma2022survey} and cybersecurity~\cite{ma2021comprehensive}. Thus, graph neural models have been proposed. For example, \textit{Graph2Vec} proposes graph embedding techniques~\cite{narayanan2017graph2vec} which utilizes the subgraph sampling and \textit{Doc2Vec} (i.e., skipgram algorithm)~\cite{le2014distributed}. Others attempt to convert graph data into transformer-compatible inputs~\cite{yang2021graphformers,kim2022pure}. However, these approaches do not fully account for locality, specifically the distance between nodes, in graphs.
To address this issue, the message passing method~\cite{gilmer2017neural} was proposed, which constructs the neural model by connecting the cells along the edges. Besides, its incremental work, graph attention networks (GATs)~\cite{velivckovic2017graph}, integrates with the attention mechanism. Here, \dupin{} uses the modified GAT to facilitate masked learning.

\section{Limitations}

\noindent\textbf{Detection Capability.}
Due to the use of supervision in learning, \dupin{} can only detect attack types that have been previously observed. As a result, it may be vulnerable to rare attacks, making its effectiveness dependent on the attack coverage of the dataset. 
For example, it is difficult to detect the sabotage attack with \dupin's model trained only with compromise phases.
Therefore, \dupin{} requires some efforts in data collection, though this is reduced by the extensive pretraining.
Our primary results may result from the some degree of similarity across attack campaigns. Our assumption is that different attacks often exhibit common, fundamental attack kill-chains, which is reflected in ATT\&CK TTP~\cite{mitre} where various attack groups employ same tactics but different procedures. Although \dupin benefits from flexibility in detecting varied attack manifestations, it cannot detect unseen attack purposes whose underlying attack fundamentals are distinct.

\noindent\textbf{Use of \dupin.} While real-time attack analysis is highly demanded across various infrastructures (e.g., cyber-physical systems), existing automation papers still implement postmortem forensics (i.e., analysis after compromise) rather than on-the-fly attack prevention. \dupin is no exception. First, the real-time responses require higher true negative rates. Second, the complexity for data pre-processing (e.g., parsing, graph constructions) needs to improve. As discussed in our case study, an analyst can only use \dupin to confirm the specific damages/losses by attack for instant responses.
However, the core idea of \dupin is general and applicable to different domains. In a closed network such as power grids, \dupin can aim for the real-time analysis because their systems exhibit periodicity, strict security rules, making our models effectively and efficiently catch the anomalies in the system logs.

\section{Conclusion}
We propose a learning-based forensic system, \dupin, which is the first to apply the workflow of (i) massive self-supervised pretraining on benign audit logs and (ii) few shot learning on attack logs as a supervised fine-tuning. We evaluate \dupin{} across 25 APT attack sets, including four different sources, demonstrating that \dupin{} outperforms one-class learners, alternative GNN architectures and ablated models.

\section{Acknowledgements}
We would like to thank the anonymous reviewers for their valuable feedback and insightful comments.
This work was funded in part by the National Science Foundation (NSF) Awards SHF-1901242, SHF-1910300, Proto-OKN 2333736, IIS-2416835, ONR N00014-23-1-2081, and Amazon. Any opinions, findings and conclusions or recommendations expressed in this material are those of the authors and do not necessarily reflect the views of the sponsors

\section{Ethical Considerations}

\noindent\textbf{Statement} Our work relies on attack datasets, which carry the risk of being misused for malicious purposes. We use the attack knowledge acquired from the dataset solely for research purposes, specifically defense research. Any subsequent work that utilizes our code, datasets, or artifacts is expected to adhere to the same code of conduct, including proper attribution and ethical use of the resources.

\noindent\textbf{Stakeholders.} First, security analysts and incident responders are the intended beneficiaries of DUPIN. They are tasked with triaging vast volumes of audit logs and stand to gain from reduced manual workload during postmortem forensic investigation. Furthermore, adversaries must also be considered. Attackers can read our paper and potentially adapt their tradecraft to evade detection, and could in principle attempt to poison the few-shot training set if they gained influence over its curation.

\noindent\textbf{Impacts on stakeholders.} For analysts, DUPIN's principal positive impact is reduction in alert-triage effort through improved true positive rates and lowered false positives relative to one-class baselines. However, false negatives may cause genuine attacks to be missed, and false positives may waste analyst time. For adversaries, publication of DUPIN provides a roadmap of what learning-based detectors now catch, which could inform evasion attempts. We judge this disclosure to be value neutral, as defenders typically benefit more from open methods than adversaries gain from studying them.

\noindent\textbf{Mitigations.} We took several steps to mitigate the identified risks. All experiments use publicly released APT datasets whose collection was conducted by their original authors with appropriate authorization. No new attack traffic was generated against live systems and no human subjects were involved. 

\noindent\textbf{Decision to conduct the research.} We weighed the dual-use nature of any detection research before beginning this work. Learning-based attack forensics has been an open and active research area at top security venues for years. We judged that \dupin would more likely benefit defenders by reducing analyst burden in real SOC environments.

\section{Open Science}
We release our source code and instructions to replicate our evaluation results. The original datasets (i.e., audit logs) need to be downloaded from their original repository. 
Additionally, we release an API toolkit for audit analysis, aiming to facilitate future work. The tool supports automated parsing from raw logs and graph constructions with a few configuration-level efforts. The link to the tool and implementation of \dupin can be found in Section~\ref{sec:intro}. The detailed information and tool demonstration are in repository.

\bibliographystyle{plain}
\bibliography{anothology}

\appendix
\section{The performance of \dupin on the relaxed cross-validations.}\label{subsec:relaxed-corss-val}

Table~\ref{table:result-relaxed-cross-val} presents the performance of \dupin under relaxed cross-validation settings, where each attack set is evaluated using a model trained on data from the same dataset source, excluding the target set itself. We follow all remaining experimental settings in accordance with the original evaluation presented in Table~\ref{table:result-main}.



\begin{table}[H]
\resizebox{\columnwidth}{!}{%
\begin{tabular}{|c|ccc|ccc|ccc|}
\hline
 &
  \multicolumn{3}{c|}{\dupin} &
  \multicolumn{3}{c|}{\textbf{AirTag}} &
  \multicolumn{3}{c|}{\textbf{MAGIC}} \\ \cline{2-10} 
\multirow{-2}{*}{ID} &
  \textbf{TPR} &
  \textbf{TNR} &
  \textbf{AUR.} &
  \textbf{TPR} &
  \textbf{TNR} &
  \textbf{AUR} &
  \textbf{TPR} &
  \textbf{TNR} &
  \textbf{AUR} \\ \hline
1 &
  0.89 &
  0.86 &
  0.92 &
  \cellcolor[HTML]{EFEFEF}1.00 &
  0.73 &
  0.85 &
  0.78 &
  0.75 &
  0.83 \\ \cline{1-1}
2 &
  \cellcolor[HTML]{EFEFEF}0.83 &
  \cellcolor[HTML]{EFEFEF}0.99 &
  \cellcolor[HTML]{EFEFEF}0.82 &
  \cellcolor[HTML]{EFEFEF}0.17 &
  \cellcolor[HTML]{EFEFEF}0.57 &
  \cellcolor[HTML]{EFEFEF}0.14 &
  \cellcolor[HTML]{EFEFEF}0.83 &
  \cellcolor[HTML]{EFEFEF}0.95 &
  \cellcolor[HTML]{EFEFEF}0.94 \\ \cline{1-1}
3 &
  1.00 &
  1.00 &
  0.92 &
  1.00 &
  0.48 &
  0.48 &
  1.00 &
  0.79 &
  0.79 \\ \cline{1-1}
4 &
  \cellcolor[HTML]{EFEFEF}1.00 &
  \cellcolor[HTML]{EFEFEF}0.98 &
  \cellcolor[HTML]{EFEFEF}0.98 &
  \cellcolor[HTML]{EFEFEF}1.00 &
  \cellcolor[HTML]{EFEFEF}1.00 &
  \cellcolor[HTML]{EFEFEF}1.00 &
  \cellcolor[HTML]{EFEFEF}1.00 &
  \cellcolor[HTML]{EFEFEF}0.94 &
  \cellcolor[HTML]{EFEFEF}0.94 \\ \cline{1-1}
5 &
  0.67 &
  1.00 &
  0.64 &
  0.67 &
  0.64 &
  0.55 &
  0.33 &
  0.42 &
  0.15 \\ \cline{1-1}
6 &
  \cellcolor[HTML]{EFEFEF}{\color[HTML]{000000} 0.75} &
  \cellcolor[HTML]{EFEFEF}{\color[HTML]{000000} 0.58} &
  \cellcolor[HTML]{EFEFEF}{\color[HTML]{000000} 0.55} &
  \cellcolor[HTML]{EFEFEF}{\color[HTML]{000000} 1.00} &
  \cellcolor[HTML]{EFEFEF}{\color[HTML]{000000} 0.20} &
  \cellcolor[HTML]{EFEFEF}{\color[HTML]{000000} 0.20} &
  \cellcolor[HTML]{EFEFEF}{\color[HTML]{000000} 1.00} &
  \cellcolor[HTML]{EFEFEF}{\color[HTML]{000000} 0.08} &
  \cellcolor[HTML]{EFEFEF}{\color[HTML]{000000} 0.15} \\ \cline{1-1}
7 &
  1.00 &
  0.71 &
  0.85 &
  0.71 &
  0.34 &
  0.34 &
  0.71 &
  0.63 &
  0.65 \\ \cline{1-1}
8 &
  \cellcolor[HTML]{EFEFEF}1.00 &
  \cellcolor[HTML]{EFEFEF}0.94 &
  \cellcolor[HTML]{EFEFEF}0.98 &
  \cellcolor[HTML]{EFEFEF}1.00 &
  \cellcolor[HTML]{EFEFEF}0.36 &
  \cellcolor[HTML]{EFEFEF}0.68 &
  \cellcolor[HTML]{EFEFEF}0.54 &
  \cellcolor[HTML]{EFEFEF}0.34 &
  \cellcolor[HTML]{EFEFEF}0.25 \\ \cline{1-1}
9 &
  1.00 &
  1.00 &
  1.00 &
  0.00 &
  1.00 &
  0.50 &
  0.75 &
  0.39 &
  0.36 \\ \cline{1-1}
10 &
  \cellcolor[HTML]{EFEFEF}0.98 &
  \cellcolor[HTML]{EFEFEF}0.94 &
  \cellcolor[HTML]{EFEFEF}0.98 &
  \cellcolor[HTML]{EFEFEF}0.92 &
  \cellcolor[HTML]{EFEFEF}0.44 &
  \cellcolor[HTML]{EFEFEF}0.43 &
  \cellcolor[HTML]{EFEFEF}0.70 &
  \cellcolor[HTML]{EFEFEF}0.53 &
  \cellcolor[HTML]{EFEFEF}0.45 \\ \cline{1-1}
11 &
  0.83 &
  0.74 &
  0.80 &
  0.00 &
  1.00 &
  0.56 &
  1.00 &
  0.00 &
  0.02 \\ \cline{1-1}
12 &
  \cellcolor[HTML]{EFEFEF}1.00 &
  \cellcolor[HTML]{EFEFEF}0.42 &
  \cellcolor[HTML]{EFEFEF}0.78 &
  \cellcolor[HTML]{EFEFEF}1.00 &
  \cellcolor[HTML]{EFEFEF}0.33 &
  \cellcolor[HTML]{EFEFEF}0.67 &
  \cellcolor[HTML]{EFEFEF}0.73 &
  \cellcolor[HTML]{EFEFEF}0.17 &
  \cellcolor[HTML]{EFEFEF}0.23 \\ \cline{1-1}
13 &
  1.00 &
  1.00 &
  1.00 &
  1.00 &
  0.45 &
  0.45 &
  0.50 &
  0.39 &
  0.27 \\ \cline{1-1}
14 &
  \cellcolor[HTML]{EFEFEF}1.00 &
  \cellcolor[HTML]{EFEFEF}0.84 &
  \cellcolor[HTML]{EFEFEF}0.83 &
  \cellcolor[HTML]{EFEFEF}1.00 &
  \cellcolor[HTML]{EFEFEF}0.68 &
  \cellcolor[HTML]{EFEFEF}0.68 &
  \cellcolor[HTML]{EFEFEF}1.00 &
  \cellcolor[HTML]{EFEFEF}0.35 &
  \cellcolor[HTML]{EFEFEF}0.35 \\ \cline{1-1}
15 &
  1.00 &
  0.74 &
  0.91 &
  1.00 &
  0.81 &
  0.85 &
  1.00 &
  0.13 &
  0.27 \\ \cline{1-1}
16 &
  \cellcolor[HTML]{EFEFEF}1.00 &
  \cellcolor[HTML]{EFEFEF}0.76 &
  \cellcolor[HTML]{EFEFEF}0.90 &
  \cellcolor[HTML]{EFEFEF}1.00 &
  \cellcolor[HTML]{EFEFEF}0.67 &
  \cellcolor[HTML]{EFEFEF}0.79 &
  \cellcolor[HTML]{EFEFEF}0.56 &
  \cellcolor[HTML]{EFEFEF}0.50 &
  \cellcolor[HTML]{EFEFEF}0.43 \\ \cline{1-1}
17 &
  1.00 &
  0.75 &
  0.92 &
  1.00 &
  0.46 &
  0.63 &
  0.55 &
  0.35 &
  0.33 \\ \cline{1-1}
18 &
  \cellcolor[HTML]{EFEFEF}0.97 &
  \cellcolor[HTML]{EFEFEF}0.98 &
  \cellcolor[HTML]{EFEFEF}0.98 &
  \cellcolor[HTML]{EFEFEF}1.00 &
  \cellcolor[HTML]{EFEFEF}0.90 &
  \cellcolor[HTML]{EFEFEF}0.95 &
  \cellcolor[HTML]{EFEFEF}0.52 &
  \cellcolor[HTML]{EFEFEF}0.39 &
  \cellcolor[HTML]{EFEFEF}0.35 \\ \cline{1-1}
19 &
  0.60 &
   &
  0.50 &
  1.00 &
   &
  0.72 &
  0.27 &
   &
  0.31 \\ \cline{1-1}
20 &
  \cellcolor[HTML]{EFEFEF}0.87 &
  \multirow{-2}{*}{0.48} &
  \cellcolor[HTML]{EFEFEF}0.63 &
  \cellcolor[HTML]{EFEFEF}0.33 &
  \multirow{-2}{*}{0.60} &
  \cellcolor[HTML]{EFEFEF}0.55 &
  \cellcolor[HTML]{EFEFEF}0.73 &
  \multirow{-2}{*}{0.77} &
  \cellcolor[HTML]{EFEFEF}0.74 \\ \cline{1-1}
21 &
  0.31 &
  \cellcolor[HTML]{EFEFEF} &
  0.24 &
  1.00 &
  \cellcolor[HTML]{EFEFEF} &
  0.96 &
  0.98 &
  \cellcolor[HTML]{EFEFEF} &
  0.95 \\ \cline{1-1}
22 &
  \cellcolor[HTML]{EFEFEF}0.69 &
  \multirow{-2}{*}{\cellcolor[HTML]{EFEFEF}0.34} &
  \cellcolor[HTML]{EFEFEF}0.44 &
  \cellcolor[HTML]{EFEFEF}0.92 &
  \multirow{-2}{*}{\cellcolor[HTML]{EFEFEF}0.92} &
  \cellcolor[HTML]{EFEFEF}0.95 &
  \cellcolor[HTML]{EFEFEF}0.58 &
  \multirow{-2}{*}{\cellcolor[HTML]{EFEFEF}0.88} &
  \cellcolor[HTML]{EFEFEF}0.77 \\ \cline{1-1}
23 &
  0.41 &
   &
  0.45 &
  0.97 &
   &
  0.94 &
  0.29 &
   &
  0.40 \\ \cline{1-1}
24 &
  \cellcolor[HTML]{EFEFEF}1.00 &
  \multirow{-2}{*}{0.51} &
  \cellcolor[HTML]{EFEFEF}0.94 &
  \cellcolor[HTML]{EFEFEF}1.00 &
  \multirow{-2}{*}{0.90} &
  \cellcolor[HTML]{EFEFEF}0.95 &
  \cellcolor[HTML]{EFEFEF}0.40 &
  \multirow{-2}{*}{0.54} &
  \cellcolor[HTML]{EFEFEF}0.48 \\ \cline{1-1}
25 &
  0.77 &
  \cellcolor[HTML]{EFEFEF}0.79 &
  0.71 &
  0.69 &
  \cellcolor[HTML]{EFEFEF}0.35 &
  0.44 &
  0.92 &
  \cellcolor[HTML]{EFEFEF}0.90 &
  0.93 \\ \hline
Avg. &
  0.86 &
  0.83 &
  0.80 &
  0.82 &
  0.74 &
  0.65 &
  0.71 &
  0.51 &
  0.55 \\ \hline
\end{tabular}
}

\caption{\centering Results of \dupin and baselines on the relaxed cross validations.}
\label{table:result-relaxed-cross-val}
\end{table}

In the relaxed cross-validations, \dupin (+13\% of TPR, +14\% of TNR) and \texttt{AirTag} (+38\% of TPR, +22\% of TNR) show increased performance, while \texttt{MAGIC} is slightly affected with similar scores (+6\% of TPR, -1\% of TNR). In particular, both \dupin and all baselines exhibit higher true positive rates (i.e., TPRs) in this training configuration. This demonstrates the impact of training/validation cross-validation setups and highlights the need for careful evaluation design.

\section{The performance of \dupin on the non-pretrained model.}\label{subsec:result-ablation}
 Out of 25 cases, 18 show increases in AUROC accuracy, with 11 of them exhibiting an AUROC improvement of even more than 20\%. The TPR of the non-pretrained model (61\%) is lower than that of the strongest baseline (65\%, \texttt{Magic}) and their difference in AUROC is slim (3\%), highlighting that extensive pretraining is a major factor contributing to the superiority of the few-shot approach. 

\begin{table}[H]
{
\centering
\small
\resizebox{\columnwidth}{!}{%
\begin{tabular}{|c|ccc|ccc|c|}
\hline
 &
  \multicolumn{3}{c|}{\textbf{\dupin w/ pretrain}} &
  \multicolumn{3}{c|}{\textbf{\dupin wo/ pretrain}} &
   \\ \cline{2-7}
\multirow{-2}{*}{\textbf{ID}} &
  \textbf{TPR} &
  \textbf{TNR} &
  \textbf{AUR.} &
  \multicolumn{1}{l}{\textbf{TPR}} &
  \multicolumn{1}{l}{\textbf{TNR}} &
  \textbf{AUR.} &
  \multirow{-2}{*}{\textbf{$\Delta$ AUR.}} \\ \hline
1 &
  \textbf{1.00} &
  0.76 &
  \textbf{0.97} &
  \textbf{1.00} &
  \textbf{0.78} &
  0.91 &
  \boldblue{+ 0.06} \\ \cline{1-1} 
2 &
  \cellcolor[HTML]{EFEFEF}\textbf{0.83} &
  \cellcolor[HTML]{EFEFEF}\textbf{0.82} &
  \cellcolor[HTML]{EFEFEF}\textbf{0.80} &
  \cellcolor[HTML]{EFEFEF}0.67 &
  \cellcolor[HTML]{EFEFEF}0.59 &
  \cellcolor[HTML]{EFEFEF}0.59 &
  \cellcolor[HTML]{EFEFEF}\boldblue{+ 0.21} \\ \cline{1-1}
3 &
  \textbf{1.00} &
  \textbf{0.76} &
  \textbf{0.99} &
  \textbf{1.00} &
  \textbf{0.76} &
  0.76 &
  \boldblue{+ 0.23} \\ \cline{1-1}
4 &
  \cellcolor[HTML]{EFEFEF}\textbf{1.00} &
  \cellcolor[HTML]{EFEFEF}\textbf{0.95} &
  \cellcolor[HTML]{EFEFEF}\textbf{0.99} &
  \cellcolor[HTML]{EFEFEF}\textbf{1.00} &
  \cellcolor[HTML]{EFEFEF}0.67 &
  \cellcolor[HTML]{EFEFEF}0.68 &
  \cellcolor[HTML]{EFEFEF}\boldblue{+ 0.31} \\ \cline{1-1}
5 &
  0.00 &
  \textbf{0.99} &
  \textbf{0.65} &
  0.00 &
  0.97 &
  0.30 &
  \boldblue{+ 0.35} \\ \cline{1-1}
6 &
  \cellcolor[HTML]{EFEFEF}0.00 &
  \cellcolor[HTML]{EFEFEF}\textbf{0.99} &
  \cellcolor[HTML]{EFEFEF}0.41 &
  \cellcolor[HTML]{EFEFEF}0.00 &
  \cellcolor[HTML]{EFEFEF}\textbf{0.99} &
  \cellcolor[HTML]{EFEFEF}0.54 &
  \cellcolor[HTML]{EFEFEF} \boldred{- 0.13} \\ \cline{1-1}
7 &
  \textbf{1.00} &
  \textbf{0.83} &
  \textbf{0.94} &
  \textbf{1.00} &
  0.19 &
  0.28 &
  \boldblue{+ 0.66} \\ \cline{1-1}
8 &
  \cellcolor[HTML]{EFEFEF}\textbf{0.85} &
  \cellcolor[HTML]{EFEFEF}\textbf{0.78} &
  \cellcolor[HTML]{EFEFEF}\textbf{0.87} &
  \cellcolor[HTML]{EFEFEF}0.54 &
  \cellcolor[HTML]{EFEFEF}0.48 &
  \cellcolor[HTML]{EFEFEF}0.49 &
  \cellcolor[HTML]{EFEFEF}\boldblue{+ 0.38} \\ \cline{1-1}
9 &
  0.25 &
  \textbf{0.68} &
  0.45 &
  \textbf{0.50} &
  0.51 &
  \textbf{0.61} &
  \boldred{- 0.16} \\ \cline{1-1}
10 &
  \cellcolor[HTML]{EFEFEF}0.68 &
  \cellcolor[HTML]{EFEFEF}\textbf{0.68} &
  \cellcolor[HTML]{EFEFEF}0.68 &
  \cellcolor[HTML]{EFEFEF}\textbf{0.76} &
  \cellcolor[HTML]{EFEFEF}0.60 &
  \cellcolor[HTML]{EFEFEF}\textbf{0.70} &
  \cellcolor[HTML]{EFEFEF}- 0.02 \\ \cline{1-1}
11 &
  \textbf{1.00} &
  \textbf{0.89} &
  \textbf{0.99} &
  \textbf{1.00} &
  \textbf{0.89} &
  \textbf{0.99} &
  0 \\ \cline{1-1}
12 &
  \cellcolor[HTML]{EFEFEF}\textbf{0.45} &
  \cellcolor[HTML]{EFEFEF}\textbf{0.85} &
  \cellcolor[HTML]{EFEFEF}\textbf{0.70} &
  \cellcolor[HTML]{EFEFEF}0.36 &
  \cellcolor[HTML]{EFEFEF}\textbf{0.85} &
  \cellcolor[HTML]{EFEFEF}0.50 &
  \cellcolor[HTML]{EFEFEF}\boldblue{+ 0.20} \\ \cline{1-1}
13 &
  \textbf{1.00} &
  \textbf{0.45} &
  \textbf{0.74} &
  0.50 &
  0.42 &
  0.50 &
  \boldblue{+ 0.24} \\ \cline{1-1}
14 &
  \cellcolor[HTML]{EFEFEF}\textbf{1.00} &
  \cellcolor[HTML]{EFEFEF}0.32 &
  \cellcolor[HTML]{EFEFEF}\textbf{0.84} &
  \cellcolor[HTML]{EFEFEF}\textbf{1.00} &
  \cellcolor[HTML]{EFEFEF}\textbf{0.45} &
  \cellcolor[HTML]{EFEFEF}0.65 &
  \cellcolor[HTML]{EFEFEF}\boldblue{+ 0.19} \\ \cline{1-1}
15 &
  0.50 &
  0.48 &
  \textbf{0.62} &
  \textbf{0.67} &
  0.55 &
  0.55 &
  \boldblue{+ 0.07} \\ \cline{1-1}
16 &
  \cellcolor[HTML]{EFEFEF}\textbf{0.78} &
  \cellcolor[HTML]{EFEFEF}\textbf{0.57} &
  \cellcolor[HTML]{EFEFEF}\textbf{0.69} &
  \cellcolor[HTML]{EFEFEF}0.33 &
  \cellcolor[HTML]{EFEFEF}0.41 &
  \cellcolor[HTML]{EFEFEF}0.31 &
  \cellcolor[HTML]{EFEFEF}\boldblue{+ 0.38} \\ \cline{1-1}
17 &
  \textbf{0.82} &
  \textbf{0.65} &
  \textbf{0.69} &
  0.36 &
  0.53 &
  0.36 &
  \boldblue{+ 0.33} \\ \cline{1-1}
18 &
  \cellcolor[HTML]{EFEFEF}0.68 &
  \cellcolor[HTML]{EFEFEF}\textbf{0.56} &
  \cellcolor[HTML]{EFEFEF}\textbf{0.66} &
  \cellcolor[HTML]{EFEFEF}\textbf{0.71} &
  \cellcolor[HTML]{EFEFEF}0.37 &
  \cellcolor[HTML]{EFEFEF}0.56 &
  \cellcolor[HTML]{EFEFEF}\boldblue{+ 0.10} \\ \cline{1-1}
19 &
  0.53 &
   &
  \textbf{0.66} &
  \textbf{0.60} &
   &
  0.63 &
  + 0.03 \\ \cline{1-1}
20 &
  \cellcolor[HTML]{EFEFEF}\textbf{0.87} &
  \multirow{-2}{*}{0.60} &
  \cellcolor[HTML]{EFEFEF}0.77 &
  \cellcolor[HTML]{EFEFEF}0.47 &
  \multirow{-2}{*}{\textbf{0.66}} &
  \cellcolor[HTML]{EFEFEF}\textbf{0.82} &
  \cellcolor[HTML]{EFEFEF}- 0.05 \\ \cline{1-1}
21 &
  \textbf{1.00} &
  \cellcolor[HTML]{EFEFEF} &
  \textbf{0.96} &
  0.99 &
  \cellcolor[HTML]{EFEFEF} &
  0.88 &
  \boldblue{+ 0.08} \\ \cline{1-1}
22 &
  \cellcolor[HTML]{EFEFEF}\textbf{0.87} &
  \multirow{-2}{*}{\cellcolor[HTML]{EFEFEF}0.56} &
  \cellcolor[HTML]{EFEFEF}0.81 &
  \cellcolor[HTML]{EFEFEF}0.82 &
  \multirow{-2}{*}{\cellcolor[HTML]{EFEFEF}\textbf{0.89}} &
  \cellcolor[HTML]{EFEFEF}\textbf{0.91} &
  \cellcolor[HTML]{EFEFEF}\boldred{- 0.10} \\ \cline{1-1}
23 &
  \textbf{0.64} &
   &
  \textbf{0.51} &
  0.07 &
   &
  0.28 &
  \boldblue{+ 0.23} \\ \cline{1-1}
24 &
  \cellcolor[HTML]{EFEFEF}0.45 &
  \multirow{-2}{*}{0.47} &
  \cellcolor[HTML]{EFEFEF}0.54 &
  \cellcolor[HTML]{EFEFEF}0.69 &
  \multirow{-2}{*}{\textbf{0.87}} &
  \cellcolor[HTML]{EFEFEF}\textbf{0.73} &
  \cellcolor[HTML]{EFEFEF}\boldred{- 0.19} \\ \cline{1-1}
25 &
  \textbf{1.00} &
  0.47 &
  \textbf{0.93} &
  0.15 &
  0.89 &
  0.58 &
  \boldblue{+ 0.32} \\ \hline
Avg. &
  \textbf{0.73} &
  \textbf{0.69} &
  \textbf{0.75} &
  0.61 &
  0.65 &
  0.59 &
  \boldblue{+ 0.16} \\ \hline
\end{tabular}
}
}
\caption{\centering Performance difference between pretrained and non-pretrained models. $\Delta$ AUR. is the performance gain ($\Delta$ AUR annotation: \boldblue{+ xx}: $\textbf{gain} > 0.05$, \boldred{-- xx}: $\textbf{gain} < -0.05$)}
\label{table:result-ablation}
\vspace{-4mm}
\end{table}

\section{Usage of Darpa Trace Dataset.}\label{subsec:usage-darpa-dataset}
Table~\ref{table:darpa-dataset-usage} shows the use of {\tt Darpa TRACE} dataset. The DARPA Trace datasets, including both E3 and E5 versions, provide a total of 9 tracepoints from {\em Linux} systems and 6 tracepoints from {\em Windows} systems. Among them, we have collected a dozen attack sets based on its official documentation. The tracepoint source for each attack set is specified in Table~\ref{table:darpa-dataset-usage}. Note that each benign set is derived from the same tracepoint as its corresponding attack set.

\begin{table}[H]{
\resizebox{\columnwidth}{!}{%
\begin{tabular}{|c|c|c|c|}
\hline
\textbf{Platform}        & \textbf{Trace Point}  & \textbf{Trace Time}             & \textbf{Used by}    \\ \hline
\multirow{9}{*}{Linux}   & Cadets 1 (E3)         & 04/02, 18:07 $\sim$ 04/06, 12:08 & Attack ID: 1        \\ \cline{2-4} 
                         & Cadets 2 (E3)         & 04/06, 14:01 $\sim$ 04/11, 15:16 & -                   \\ \cline{2-4} 
                         & Cadets 3 (E3)         & 04/11, 16:36 $\sim$ 04/13, 17:35 & Attack ID: 2, 3     \\ \cline{2-4} 
                         & Trace 1 (E3)          & 04/02, 17:14 $\sim$ 04/13, 09:06 & -                   \\ \cline{2-4} 
                         & Trace 2 (E3)          & 04/13, 09:56 $\sim$ 04/13, 17:09 & Attack ID: 4, 5     \\ \cline{2-4} 
                         & Cadets 1 (E5)         & 05/07, 08:44 $\sim$ 05/17, 17:13 & Attack ID: 7        \\ \cline{2-4} 
                         & Trace 1 (E5)          & 05/07, 09:15 $\sim$ 05/17, 17:44 & -                   \\ \cline{2-4} 
                         & Trace 2 (E5)          & 05/07, 09:15 $\sim$ 05/17, 16:07 & -                   \\ \cline{2-4} 
                         & Trace 3 (E5)          & 05/07, 10:35 $\sim$ 05/17, 17:42 & Attack ID: 6        \\ \hline
\multirow{6}{*}{Windows} & Fivedirections 1 (E3) & 04/03, 08:21 $\sim$ 04/04, 08:42 & -                   \\ \cline{2-4} 
                         & Fivedirections 2 (E3) & 04/04, 09:53 $\sim$ 04/12, 23:28 & Attack ID: 8, 9, 10 \\ \cline{2-4} 
                         & Fivedirections 3 (E3) & 04/13, 09:37 $\sim$ 04/13, 17:08 & -                   \\ \cline{2-4} 
                         & Fivedirections 1 (E5) & 05/07, 09:02 $\sim$ 05/17, 17:36 & Attack ID: 11       \\ \cline{2-4} 
                         & Fivedirections 2 (E5) & 05/07, 09:05 $\sim$ 05/17, 17:14 & Attack ID: 12       \\ \cline{2-4} 
                         & Fivedirections 3 (E5) & 05/07, 09:08 $\sim$ 05/17, 17:38 & -                   \\ \hline
\end{tabular} %
}
}
\caption{\centering Usage of Darpa TRACE dataset.}
\label{table:darpa-dataset-usage}
\end{table}

Among them, the DARPA Trace dataset includes 5 tracepoints for {\em Linux} and 6 tracepoints for {\em Windows}, each with a tracing span of over one week. We use these tracepoints for pretraining \dupin. During pretraining, we exclude any trace durations that overlap with the periods used for constructing the attack and benign sets to ensure strict separation.

\section{Subdivided Results in Our Evaluations}\label{subsec:full-flash-kairos-gin-gcn}

This section holds the full results of Table~\ref{table:comparison-flash-kairos},~\ref{table:abalation-no-attentive},~\ref{table:dupin-vs-velox}.
\begin{table}[H]
{
\centering
\scriptsize
\resizebox{0.9\columnwidth}{!}{%
{
\begin{tabular}{|c|cccccc|}
\hline
 &
  \multicolumn{3}{c|}{\texttt{Flash}} &
  \multicolumn{3}{c|}{\texttt{Kairos}} \\ \cline{2-7} 
  \multirow{-2}{*}{ID} & 
  \textbf{TPR} &
  \textbf{TNR} &
  \textbf{AUR.} &
  \textbf{TPR} &
  \textbf{TNR} &
  \textbf{AUR.}  \\ \hline
1 &
  {0.78} &
  {0.53} &
  {0.69} &
  {0.67} &
  {0.39} &
  {0.54} \\ \cline{1-1}
2 &
  \cellcolor[HTML]{EFEFEF}{0.83} &
  \cellcolor[HTML]{EFEFEF}{0.39} & 
  \cellcolor[HTML]{EFEFEF}{0.61} &
  \cellcolor[HTML]{EFEFEF}{0.33} &
  \cellcolor[HTML]{EFEFEF}{0.42} & 
  \cellcolor[HTML]{EFEFEF}{0.20} \\ \cline{1-1}
3 &
  {0.00} &
  {0.52} &
  {0.35} &
  {1.00} &
  {0.30} &
  {0.20}  \\ \cline{1-1}
4 &
  \cellcolor[HTML]{EFEFEF}{0.76} &
  \cellcolor[HTML]{EFEFEF}0.68 &
  \cellcolor[HTML]{EFEFEF}0.76 &
  \cellcolor[HTML]{EFEFEF}{1.00} &
  \cellcolor[HTML]{EFEFEF}0.22 &
  \cellcolor[HTML]{EFEFEF}0.63  \\ \cline{1-1}
5 &
  {1.00} &
  {0.57} &
  {0.88} &
  {0.86} &
  {0.49} &
  {0.86}  \\ \cline{1-1}
6 &
  \cellcolor[HTML]{EFEFEF}0.00 &
  \cellcolor[HTML]{EFEFEF}{1.00} &
  \cellcolor[HTML]{EFEFEF}{0.13} &
  \cellcolor[HTML]{EFEFEF}0.00 &
  \cellcolor[HTML]{EFEFEF}{1.00} &
  \cellcolor[HTML]{EFEFEF}{1.00}  \\ \cline{1-1}
7 &
  {1.00} &
  {0.47} &
  {0.81} &
  {0.57} &
  {0.06} &
  {0.07}  \\ \cline{1-1}
8 &
  \cellcolor[HTML]{EFEFEF}{0.77} &
  \cellcolor[HTML]{EFEFEF}{0.28} &
  \cellcolor[HTML]{EFEFEF}{0.47} &
  \cellcolor[HTML]{EFEFEF}{1.00} &
  \cellcolor[HTML]{EFEFEF}{0.42} &
  \cellcolor[HTML]{EFEFEF}{0.77} \\ \cline{1-1}
9 &
  {0.50} &
  {0.25} &
  {0.77} &
  {0.00} &
  {0.35} &
  {0.30}  \\ \cline{1-1}
10 &
  \cellcolor[HTML]{EFEFEF}{0.44} &
  \cellcolor[HTML]{EFEFEF}{0.13} &
  \cellcolor[HTML]{EFEFEF}{0.81} &
  \cellcolor[HTML]{EFEFEF}{0.96} &
  \cellcolor[HTML]{EFEFEF}{0.17} &
  \cellcolor[HTML]{EFEFEF}{0.49}  \\ \cline{1-1}
11 &
  1.00 &
  {0.01} &
  {0.81} &
  1.00 &
  {0.91} &
  {0.95}  \\ \cline{1-1}
12 &
  \cellcolor[HTML]{EFEFEF}{1.00} &
  \cellcolor[HTML]{EFEFEF}{0.24} &
  \cellcolor[HTML]{EFEFEF}{0.54} &
  \cellcolor[HTML]{EFEFEF}{1.00} &
  \cellcolor[HTML]{EFEFEF}{0.80} &
  \cellcolor[HTML]{EFEFEF}{0.96}  \\ \cline{1-1}
13 &
  {0.50} &
  {0.74} &
  {0.41} &
  {0.00} &
  {1.00} &
  {0.18}  \\ \cline{1-1}
14 &
  \cellcolor[HTML]{EFEFEF}{0.00} &
  \cellcolor[HTML]{EFEFEF}{0.77} &
  \cellcolor[HTML]{EFEFEF}{0.41} &
  \cellcolor[HTML]{EFEFEF}{0.00} &
  \cellcolor[HTML]{EFEFEF}{1.00} &
  \cellcolor[HTML]{EFEFEF}{0.18}  \\ \cline{1-1}
15 &
  {0.17} &
  0.75 &
  {0.36} &
  {0.17} &
  0.31 &
  {0.15}  \\ \cline{1-1}
16 &
  \cellcolor[HTML]{EFEFEF}{0.44} &
  \cellcolor[HTML]{EFEFEF}{0.78} &
  \cellcolor[HTML]{EFEFEF}{0.52} &
  \cellcolor[HTML]{EFEFEF}{0.00} &
  \cellcolor[HTML]{EFEFEF}{1.00} &
  \cellcolor[HTML]{EFEFEF}{0.05}  \\ \cline{1-1}
17 &
  {0.56} &
  {0.75} &
  {0.54} &
  {0.60} &
  {0.56} &
  {0.60}  \\ \cline{1-1}
18 &
  \cellcolor[HTML]{EFEFEF}0.45 &
  \cellcolor[HTML]{EFEFEF}{0.60} &
  \cellcolor[HTML]{EFEFEF}{0.46} &
  \cellcolor[HTML]{EFEFEF}0.35 &
  \cellcolor[HTML]{EFEFEF}{0.38} &
  \cellcolor[HTML]{EFEFEF}{0.32}  \\ \cline{1-1}
19 &
  {0.64} &
  &
  {0.73} &
  {1.91} &
  &
  {0.55}  \\ \cline{1-1}
20 &
  \cellcolor[HTML]{EFEFEF}{1.00} &
  \multirow{-2}{*}{0.33} &
  \cellcolor[HTML]{EFEFEF}{0.48} &
  \cellcolor[HTML]{EFEFEF}{0.83} &
  \multirow{-2}{*}{0.33} &
  \cellcolor[HTML]{EFEFEF}{0.41}  \\ \cline{1-1}
21 &
  {0.63} &
  \cellcolor[HTML]{EFEFEF} &
  {0.44} &
  {0.88} &
  \cellcolor[HTML]{EFEFEF} &
  {0.55}  \\ \cline{1-1}
22 &
  \cellcolor[HTML]{EFEFEF}0.78 &
  \multirow{-2}{*}{\cellcolor[HTML]{EFEFEF}{0.37}} &
  \cellcolor[HTML]{EFEFEF}{0.64} &
  \cellcolor[HTML]{EFEFEF}0.54 &
  \multirow{-2}{*}{\cellcolor[HTML]{EFEFEF}{0.44}} &
  \cellcolor[HTML]{EFEFEF}{0.47}  \\ \cline{1-1}
23 &
  {0.45} &
  &
  {0.39} &
  {0.56} &
  &
  {0.43}  \\ \cline{1-1}
24 &
  \cellcolor[HTML]{EFEFEF}0.53 &
  \multirow{-2}{*}{0.43} &
  \cellcolor[HTML]{EFEFEF}{0.50} &
  \cellcolor[HTML]{EFEFEF}0.35 &
  \multirow{-2}{*}{0.41} &
  \cellcolor[HTML]{EFEFEF}{0.29}  \\ \cline{1-1}
25 &
  0.54 &
  \cellcolor[HTML]{EFEFEF}{0.55} &
  {0.58} &
  0.69 &
  \cellcolor[HTML]{EFEFEF}{0.39} &
  {0.58}  \\ \cline{1-1} \hline
Avg. &
  {0.60} &
  {0.51} &
  {0.55} &
  {0.57} &
  {0.52} &
  {0.51} \\ \hline
\end{tabular} %
}
}
}
\caption{\centering Full Results on {\tt Flash} and {\tt Kairos} models.}
\label{table:result-ablation-gin}
\end{table}
\begin{table}[H]
{
\centering
\scriptsize
\resizebox{0.9\columnwidth}{!}{%
{
\begin{tabular}{|c|cccccc|}
\hline
 &
  \multicolumn{3}{c|}{\texttt{GIN}} &
  \multicolumn{3}{c|}{\texttt{GCN}} \\ \cline{2-7} 
  \multirow{-2}{*}{ID} & 
  \textbf{TPR} &
  \textbf{TNR} &
  \textbf{AUR.} &
  \textbf{TPR} &
  \textbf{TNR} &
  \textbf{AUR.}  \\ \hline
1 &
  {0.11} &
  {0.52} &
  {0.33} &
  {1.00} &
  {0.28} &
  {0.70} \\ \cline{1-1}
2 &
  \cellcolor[HTML]{EFEFEF}{0.50} &
  \cellcolor[HTML]{EFEFEF}{0.55} & 
  \cellcolor[HTML]{EFEFEF}{0.61} &
  \cellcolor[HTML]{EFEFEF}{0.83} &
  \cellcolor[HTML]{EFEFEF}{0.23} & 
  \cellcolor[HTML]{EFEFEF}{0.64} \\ \cline{1-1}
3 &
  {0.00} &
  {0.46} &
  {0.20} &
  {1.00} &
  {0.30} &
  {0.89}  \\ \cline{1-1}
4 &
  \cellcolor[HTML]{EFEFEF}{1.00} &
  \cellcolor[HTML]{EFEFEF}0.70 &
  \cellcolor[HTML]{EFEFEF}0.84 &
  \cellcolor[HTML]{EFEFEF}{0.55} &
  \cellcolor[HTML]{EFEFEF}0.77 &
  \cellcolor[HTML]{EFEFEF}0.65  \\ \cline{1-1}
5 &
  {0.67} &
  {0.59} &
  {0.61} &
  {0.00} &
  {0.73} &
  {0.49}  \\ \cline{1-1}
6 &
  \cellcolor[HTML]{EFEFEF}0.25 &
  \cellcolor[HTML]{EFEFEF}{0.77} &
  \cellcolor[HTML]{EFEFEF}{0.69} &
  \cellcolor[HTML]{EFEFEF}0.25 &
  \cellcolor[HTML]{EFEFEF}{0.87} &
  \cellcolor[HTML]{EFEFEF}{0.60}  \\ \cline{1-1}
7 &
  {1.00} &
  {0.28} &
  {0.48} &
  {0.43} &
  {0.31} &
  {0.26}  \\ \cline{1-1}
8 &
  \cellcolor[HTML]{EFEFEF}{0.54} &
  \cellcolor[HTML]{EFEFEF}{0.86} &
  \cellcolor[HTML]{EFEFEF}{0.57} &
  \cellcolor[HTML]{EFEFEF}{0.46} &
  \cellcolor[HTML]{EFEFEF}{0.79} &
  \cellcolor[HTML]{EFEFEF}{0.80} \\ \cline{1-1}
9 &
  {0.00} &
  {0.79} &
  {0.01} &
  {1.00} &
  {0.80} &
  {0.92}  \\ \cline{1-1}
10 &
  \cellcolor[HTML]{EFEFEF}{0.10} &
  \cellcolor[HTML]{EFEFEF}{0.78} &
  \cellcolor[HTML]{EFEFEF}{0.15} &
  \cellcolor[HTML]{EFEFEF}{0.82} &
  \cellcolor[HTML]{EFEFEF}{0.79} &
  \cellcolor[HTML]{EFEFEF}{0.80}  \\ \cline{1-1}
11 &
  0.00 &
  {0.82} &
  {0.62} &
  0.83 &
  {0.54} &
  {0.66}  \\ \cline{1-1}
12 &
  \cellcolor[HTML]{EFEFEF}{0.45} &
  \cellcolor[HTML]{EFEFEF}{0.75} &
  \cellcolor[HTML]{EFEFEF}{0.66} &
  \cellcolor[HTML]{EFEFEF}{0.45} &
  \cellcolor[HTML]{EFEFEF}{0.64} &
  \cellcolor[HTML]{EFEFEF}{0.49}  \\ \cline{1-1}
13 &
  {1.00} &
  {0.61} &
  {0.89} &
  {1.00} &
  {0.67} &
  {0.76}  \\ \cline{1-1}
14 &
  \cellcolor[HTML]{EFEFEF}{1.00} &
  \cellcolor[HTML]{EFEFEF}{0.55} &
  \cellcolor[HTML]{EFEFEF}{0.84} &
  \cellcolor[HTML]{EFEFEF}{1.00} &
  \cellcolor[HTML]{EFEFEF}{0.74} &
  \cellcolor[HTML]{EFEFEF}{0.82}  \\ \cline{1-1}
15 &
  {0.58} &
  0.56 &
  {0.59} &
  {1.00} &
  0.71 &
  {0.89}  \\ \cline{1-1}
16 &
  \cellcolor[HTML]{EFEFEF}{0.33} &
  \cellcolor[HTML]{EFEFEF}{0.48} &
  \cellcolor[HTML]{EFEFEF}{0.29} &
  \cellcolor[HTML]{EFEFEF}{0.67} &
  \cellcolor[HTML]{EFEFEF}{0.70} &
  \cellcolor[HTML]{EFEFEF}{0.74}  \\ \cline{1-1}
17 &
  {0.45} &
  {0.53} &
  {0.42} &
  {0.91} &
  {0.60} &
  {0.82}  \\ \cline{1-1}
18 &
  \cellcolor[HTML]{EFEFEF}0.58 &
  \cellcolor[HTML]{EFEFEF}{0.56} &
  \cellcolor[HTML]{EFEFEF}{0.59} &
  \cellcolor[HTML]{EFEFEF}0.61 &
  \cellcolor[HTML]{EFEFEF}{0.68} &
  \cellcolor[HTML]{EFEFEF}{0.74}  \\ \cline{1-1}
19 &
  {0.93} &
  &
  {0.85} &
  {1.00} &
  &
  {0.91}  \\ \cline{1-1}
20 &
  \cellcolor[HTML]{EFEFEF}{0.47} &
  \multirow{-2}{*}{0.76} &
  \cellcolor[HTML]{EFEFEF}{0.74} &
  \cellcolor[HTML]{EFEFEF}{1.00} &
  \multirow{-2}{*}{0.25} &
  \cellcolor[HTML]{EFEFEF}{0.87}  \\ \cline{1-1}
21 &
  {0.97} &
  \cellcolor[HTML]{EFEFEF} &
  {0.98} &
  {1.00} &
  \cellcolor[HTML]{EFEFEF} &
  {0.97}  \\ \cline{1-1}
22 &
  \cellcolor[HTML]{EFEFEF}0.73 &
  \multirow{-2}{*}{\cellcolor[HTML]{EFEFEF}{0.81}} &
  \cellcolor[HTML]{EFEFEF}{0.98} &
  \cellcolor[HTML]{EFEFEF}0.97 &
  \multirow{-2}{*}{\cellcolor[HTML]{EFEFEF}{0.36}} &
  \cellcolor[HTML]{EFEFEF}{0.86}  \\ \cline{1-1}
23 &
  {0.00} &
  &
  {0.65} &
  {0.69} &
  &
  {0.30}  \\ \cline{1-1}
24 &
  \cellcolor[HTML]{EFEFEF}0.16 &
  \multirow{-2}{*}{0.65} &
  \cellcolor[HTML]{EFEFEF}{0.32} &
  \cellcolor[HTML]{EFEFEF}1.00 &
  \multirow{-2}{*}{0.24} &
  \cellcolor[HTML]{EFEFEF}{0.44}  \\ \cline{1-1}
25 &
  0.92 &
  \cellcolor[HTML]{EFEFEF}{0.68} &
  {0.90} &
  1.00 &
  \cellcolor[HTML]{EFEFEF}{0.24} &
  {0.97}  \\ \cline{1-1} \hline
Avg. &
  {0.51} &
  {0.64} &
  {0.57} &
  {0.77} &
  {0.56} &
  {0.70} \\ \hline
\end{tabular} %
}
}
}
\caption{\centering Full Results on {\tt GIN} and {\tt GCN} models.}
\label{table:result-ablation-flash-kairos}
\end{table}
\begin{table}[H]
{
\centering
\scriptsize
\resizebox{.9\columnwidth}{!}{%
\begin{tabular}{|c|ccc|ccc|}
\hline
 &
  \multicolumn{3}{c|}{\texttt{TAPAS}} &
  \multicolumn{3}{c|}{\texttt{VELOX}} 
   \\ \cline{2-7}
\multirow{-2}{*}{\textbf{ID}} &
  \textbf{TPR} &
  \textbf{TNR} &
  \textbf{AUR.} &
  \multicolumn{1}{l}{\textbf{TPR}} &
  \multicolumn{1}{l}{\textbf{TNR}} &
  \textbf{AUR.}  \\ \hline
1 &
  0.11 &
  0.72 &
  0.35 &
  0.44 &
  0.86 &
  0.76\\ \cline{1-1} 
2 &
  \cellcolor[HTML]{EFEFEF}0.50 &
  \cellcolor[HTML]{EFEFEF}0.43 &
  \cellcolor[HTML]{EFEFEF}0.37 &
  \cellcolor[HTML]{EFEFEF}0.00 &
  \cellcolor[HTML]{EFEFEF}0.83 &
  \cellcolor[HTML]{EFEFEF}0.10 \\ \cline{1-1}
3 &
  0.00 &
  0.67 &
  0.02 &
  0.00 &
  0.85 &
  0.76 \\ \cline{1-1}
4 &
  \cellcolor[HTML]{EFEFEF}1.00 &
  \cellcolor[HTML]{EFEFEF}1.00 &
  \cellcolor[HTML]{EFEFEF}1.00 &
  \cellcolor[HTML]{EFEFEF}1.00 &
  \cellcolor[HTML]{EFEFEF}0.58 &
  \cellcolor[HTML]{EFEFEF}0.95  \\ \cline{1-1}
5 &
  0.00 &
  1.00 &
  0.19 &
  1.00 &
  0.62 &
  0.82  \\ \cline{1-1}
6 &
  \cellcolor[HTML]{EFEFEF}0.00 &
  \cellcolor[HTML]{EFEFEF}0.50 &
  \cellcolor[HTML]{EFEFEF}0.06 &
  \cellcolor[HTML]{EFEFEF}0.50 &
  \cellcolor[HTML]{EFEFEF}0.11 &
  \cellcolor[HTML]{EFEFEF}0.52 \\ \cline{1-1}
7 &
  0.86 &
  0.14 &
  0.26 &
  0.00 &
  0.83 &
  0.46 \\ \cline{1-1}
8 &
  \cellcolor[HTML]{EFEFEF}1.00 &
  \cellcolor[HTML]{EFEFEF}0.12 &
  \cellcolor[HTML]{EFEFEF}0.42 &
  \cellcolor[HTML]{EFEFEF}0.92 &
  \cellcolor[HTML]{EFEFEF}0.45 &
  \cellcolor[HTML]{EFEFEF}0.56  \\ \cline{1-1}
9 &
  0.75 &
  0.24 &
  0.28 &
  0.00 &
  0.43 &
  0.36 \\ \cline{1-1}
10 &
  \cellcolor[HTML]{EFEFEF}1.00 &
  \cellcolor[HTML]{EFEFEF}0.03 &
  \cellcolor[HTML]{EFEFEF}0.32 &
  \cellcolor[HTML]{EFEFEF}0.98 &
  \cellcolor[HTML]{EFEFEF}0.29 &
  \cellcolor[HTML]{EFEFEF}0.39 \\ \cline{1-1}
11 &
  1.00 &
  1.00 &
  0.38 &
  1.00 &
  0.94 &
  0.96 \\ \cline{1-1}
12 &
  \cellcolor[HTML]{EFEFEF}0.00 &
  \cellcolor[HTML]{EFEFEF}0.00 &
  \cellcolor[HTML]{EFEFEF}0.95 &
  \cellcolor[HTML]{EFEFEF}1.00 &
  \cellcolor[HTML]{EFEFEF}0.85 &
  \cellcolor[HTML]{EFEFEF}0.90 \\ \cline{1-1}
13 &
  1.00 &
  0.97 &
  0.95 &
  1.00 &
  0.58 &
  0.74  \\ \cline{1-1}
14 &
  \cellcolor[HTML]{EFEFEF}1.00 &
  \cellcolor[HTML]{EFEFEF}1.00 &
  \cellcolor[HTML]{EFEFEF}1.00 &
  \cellcolor[HTML]{EFEFEF}0.00 &
  \cellcolor[HTML]{EFEFEF}0.64 &
  \cellcolor[HTML]{EFEFEF}0.51 \\ \cline{1-1}
15 &
  0.00 &
  0.30 &
  0.87 &
  0.33 &
  0.61 &
  0.53  \\ \cline{1-1}
16 &
  \cellcolor[HTML]{EFEFEF}0.44 &
  \cellcolor[HTML]{EFEFEF}0.40 &
  \cellcolor[HTML]{EFEFEF}0.40 &
  \cellcolor[HTML]{EFEFEF}0.78 &
  \cellcolor[HTML]{EFEFEF}0.63 &
  \cellcolor[HTML]{EFEFEF}0.66 \\ \cline{1-1}
17 &
  0.56 &
  0.58 &
  0.55 &
  0.33 &
  0.55 &
  0.34  \\ \cline{1-1}
18 &
  \cellcolor[HTML]{EFEFEF}0.65 &
  \cellcolor[HTML]{EFEFEF}0.42 &
  \cellcolor[HTML]{EFEFEF}0.49 &
  \cellcolor[HTML]{EFEFEF}0.48 &
  \cellcolor[HTML]{EFEFEF}0.63 &
  \cellcolor[HTML]{EFEFEF}0.54  \\ \cline{1-1}
19 &
  0.73 &
   &
  0.46 &
  0.18 &
   &
  0.32  \\ \cline{1-1}
20 &
  \cellcolor[HTML]{EFEFEF}1.00 &
  \multirow{-2}{*}{0.16} &
  \cellcolor[HTML]{EFEFEF}0.61 &
  \cellcolor[HTML]{EFEFEF}0.00 &
  \multirow{-2}{*}{0.53} &
  \cellcolor[HTML]{EFEFEF}0.26 \\ \cline{1-1}
21 &
  0.98 &
  \cellcolor[HTML]{EFEFEF} &
  0.87 &
  0.76 &
  \cellcolor[HTML]{EFEFEF} &
  0.66 \\ \cline{1-1}
22 &
  \cellcolor[HTML]{EFEFEF}0.33 &
  \multirow{-2}{*}{\cellcolor[HTML]{EFEFEF}0.33} &
  \cellcolor[HTML]{EFEFEF}0.28 &
  \cellcolor[HTML]{EFEFEF}0.49 &
  \multirow{-2}{*}{\cellcolor[HTML]{EFEFEF}0.46} &
  \cellcolor[HTML]{EFEFEF}0.62  \\ \cline{1-1}
23 &
  0.86 &
   &
  0.76 &
  0.66 &
   &
  0.53 \\ \cline{1-1}
24 &
  \cellcolor[HTML]{EFEFEF}0.66 &
  \multirow{-2}{*}{0.55} &
  \cellcolor[HTML]{EFEFEF}0.68 &
  \cellcolor[HTML]{EFEFEF}0.82 &
  \multirow{-2}{*}{0.47} &
  \cellcolor[HTML]{EFEFEF}0.67  \\ \cline{1-1}
25 &
  0.08 &
  0.70 &
  0.24 &
  0.69 &
  0.53 &
  0.68  \\ \hline
Avg. &
  0.54 &
  0.51 &
  0.51 &
  0.54 &
  0.60 &
  0.58 \\ \hline
\end{tabular}
}
}
\caption{\centering Full Results on \texttt{VELOX} and \texttt{TAPAS}.}
\label{table:result-appendix-velox}
\vspace{-4mm}
\end{table}

\section{Subgraph Discovery Algorithm}\label{subsec:subgraph-discovery}
The algorithm takes a provenance graph and generates subgraphs. As neighboring node is randomly selected (at line 20) and added (at line 23) at each step, the subgraph grows. If the algorithm fails to find nodes with unseen neighbors more than $\gamma$ times (line 24 - 29), the search is halted, preventing the infinite loop. We also stop searching when the size of subgraph reaches $\gamma$ (at line 19).
We use $\gamma = 40$, $\delta = 100$ for the discovery of subgraphs. For complete coverage, we iterate through all nodes in the audit logs and execute the discovery graph with every node.

It is noteworthy that attack and benign subgraphs are only labeled afterward. Therefore, it complies with our experimental principle that any ground-truth information is not used during the graph discovery phase. 

\MakeRobust{\Call}
\definecolor{codeblue}{rgb}{0.01, 0.28, 0.49}
\begin{algorithm}[H]
\caption{Random Subgraph Discovery Algorithm}
\label{alg:subgraph-discovery}
\begin{algorithmic}[1]
\Function{RandomSubgraphDiscovery}{}
    \State \textcolor{codeblue}{/* $\mathcal{N},\:\mathcal{E} \text{ contain the whole provenance} $ */}
    \State \textbf{Input:} $\mathcal{N}: \text{Set of Nodes},\:\mathcal{E}: \text{Set of Edges}$
    \State \textbf{Output:} $\mathcal{S}: \text{Set of Discovered Subgraphs}$
    \ForAll{$n_s \in \mathcal{N}$}
        \State $g \gets \Call{SingleGraphDiscovery}{\mathcal{N},\;\mathcal{E}, n_s}$
        \If{$g \neq \texttt{FAIL}$}
            \State $\mathcal{S} = \mathcal{S} \cup \{g\}$ 
        \EndIf
    \EndFor    
    \State \Call{Return}{$\mathcal{S}$}
\EndFunction

\vspace{1em}

\State \textcolor{codeblue}{/* $\mathcal{N},\:\mathcal{E}$ from \Call{RandomSubgraphDiscovery}{} */}
\Function{SingleGrpahDiscovery}{}
    \State \textbf{Input:} $\mathcal{N},\:\mathcal{E},\:n_s: \text{A node to start with}$
    \State \textbf{Output:} $g: \text{The discovered subgraph}$
    \State $\texttt{limit} = 0$
    \State $g$ = $\{n_s\}$
    \While{$\Call{Size}{g} < \gamma$}
        \State $n_t \gets \Call{RandomSelect}{g} $
        \If{$\Call{Neighbors}{n_t,\:\mathcal{E}} \neq \phi$}
            \State $s = \Call{Neighbors}{n_t,\:\mathcal{N},\:\mathcal{E}}$
            \State $g = g \quad \cup $
            \State $\quad \{\Call{RandomSelect}{s} \}$
        \Else
            \State \textcolor{codeblue}{/* Limiting the search by threshold */}
            \State $\texttt{limit} = \texttt{limit} + 1$
            \If{$\texttt{limit} == \delta$}
                \State break;
            \EndIf
        \EndIf
    \EndWhile
    \State \textcolor{codeblue}{/* The size of $g$ is always greater than equal to $\gamma$ */}
    \State \Call{Return}{\texttt{g}}
\EndFunction
\end{algorithmic}
\end{algorithm}

\end{document}